\documentclass[11pt]{article}
\usepackage{amsmath,amssymb}
\usepackage{graphicx}

\usepackage[sc,small]{caption}

\newcommand{\be}{\begin{eqnarray}}
\newcommand{\ee}{\end{eqnarray}}
\newcommand{\tht}{\vartheta}

\newcommand{\A}{{\mathcal A}}
\newcommand{\K}{{\mathcal K}}
\newcommand{\M}{{\mathcal M}}
\newcommand{\N}{{\mathcal N}}
\newcommand{\R}{{\mathcal R}}

\newcommand{\thbw}[2]{\vartheta[\genfrac{}{}{0pt}{}{#1}{#2} ]}

\newcommand{\thba}[2]{\vartheta[\!\!\begin{array}{c}{\phantom{}\vspace{-0.5mm}\scriptstyle#1}%
                        \\[-1.8mm]{\scriptstyle #2}\end{array}\!\!]}

\newcommand{\beqn}{\begin{eqnarray}}
\newcommand{\eeqn}{\end{eqnarray}}
\newcommand{\tr}{{\rm tr}}

\newcommand{\ca}{{\cal A}}

\newcommand{\ci}{{\cal I}}

\newcommand{\cm}{{\cal M}}

\newcommand{\cR}{{\cal R}}

\newcommand{\ct}{{\cal T}}

\newcommand{\cz}{{\cal Z}}
\newcommand{\Ref}[1]{(\ref{#1})}

\newcommand{\zba}[2]{[\!\!\begin{array}{c}{\scriptstyle#1}%
                        \\[-1.6mm]{\scriptstyle #2}\end{array}\!\!]}
\newcommand{\twist}{\Theta}
\newcommand{\alphai}{i}

\begin{document}


\hfill LMU-ASC 81/11\\
\vspace{1cm}

\begin{center}
{\bf\LARGE
One-loop K\"ahler Metric of 
D-branes at Angles \\  }

\vspace{2.5cm}

{\large
{\bf Marcus Berg$^{+}$, }
{\bf Michael Haack$^{\dag}$, }
{\bf Jin U Kang$^{\star}$}
\vspace{1cm}

{\it
$^{+}$
Department of Physics, 
Karlstad University \\  651 88 Karlstad, Sweden\\
and\\
 Oskar Klein Center, Stockholm University\\
 Albanova University Center\\ 106 91 Stockholm, Sweden\\ [5mm]

$^{\dag}$ 
 Arnold Sommerfeld Center for Theoretical Physics \\ 
Ludwig-Maximilians-Universit\"at M\"unchen \\ 
Theresienstrasse 37, 80333 M\"unchen, Germany\\ [5mm] 
$^{\star}$ 
Department of Physics, Kim Il Sung University\\ Pyongyang, DPR  Korea
}
}

\end{center}
\vspace{8mm}

\begin{center}
{\bf Abstract}\\
\end{center}
We evaluate 
string one-loop contributions
to the K\"ahler metric of D-brane moduli 
(positions and Wilson lines), in toroidal orientifolds
with branes at angles.
Contributions
due to bulk states in the loop are known, so
 we focus on the 
contributions due to states localized at intersections of orientifold
images.
We show that these quantum corrections vanish. This does not follow from the usual
nonrenormalization theorems of supersymmetric field theory.
\clearpage
 \tableofcontents

\section{Introduction}
Since the beginning, D-branes have been very important in the formal development of string theory 
as well as in attempts to apply string theory to particle phenomenology and cosmology. 
In all three endeavors (formal, phenomenogical, cosmological), a central role is played by the D-brane moduli. These moduli fields in general include scalar fields whose vacuum expectation values
specify the location of the D-brane, as well as background values of gauge potentials (Wilson lines) 
for the gauge fields that are localized on the D-branes. The metric on the field space of these moduli, in its various incarnations, 
contributes to determining the dynamics and ground state of the theory. 

For phenomenology, although these moduli fall in the adjoint representation of the D-brane gauge group and are usually not thought of as matter fields, they have  been used as toy models for matter fields  (see e.g.\ \cite{Kachru:2007xp}). For cosmology, although these moduli are in principle charged, they are uncharged under the relevant remnant gauge group, so they are still appropriate for inflation
(see e.g.\ \cite{Baumann:2009ni} for a recent review). In general, the metric on the space of D-brane moduli is of great interest also for applications. 

Sometimes, quantum corrections to this metric can become relevant. Symmetries can render the moduli potential particularly simple or in exceptional cases, can make it vanish. Then, quantum corrections to the metric on field space could  contribute interesting dependence on the D-brane moduli. Even if the quantum corrections are not leading, they can have useful interpretations. For example, when the moduli are all fixed,
the corrections may reduce to anomalous dimensions for the D-brane moduli fields (see e.g.\  \cite{Bain:2000fb} 
for a related orientifold example), that induce scale dependence in the low-energy effective theory. 
Calculations of similar type can also be used to compute masses
for adjoint fields, that have been argued to provide one-loop Dirac gaugino masses 
(first calculated in \cite{Antoniadis:2006eb}). The virtue of these for phenomenology 
has been emphasized for example in \cite{Benakli:2009mk}.

In this paper, we continue developing the formalism for computing quantum corrections to the metric of D-brane moduli, in Type IIA
orientifolds with D6-branes at angles.  The D-brane moduli can a priori couple to orbifold-charged\footnote{This is
a charge under the group of space rotations, so should not be confused with their charge under the gauge group. They are sometimes called ``twisted'' though
we prefer to reserve this terminology for states that have non-integer operator product expansions.} open string states localized at intersections of various orbifold images of a given D-brane. In other words, a D-brane at angles can intersect its own orbifold image, and therefore also states localized at these intersections can run in an open string loop.  These corrections are referred to as $\N=1$ corrections, as they only appear
for D-branes that have nonvanishing intersection angle along all three two-tori. 

There are also loop contributions due to states not localized at intersections, when there exist branes that are parallel along one two-torus. (This is the only other nontrivial option, because branes cannot be parallel along two two-tori but not the third and still preserve supersymmetry.) Such special configurations preserve enhanced ${\cal N}=2$ supersymmetry, so those corrections are referred to as $\N=2$ corrections. The enhanced supersymmetry actually prevents string oscillators from contributing, so these corrections only come from zero modes (winding or Kaluza-Klein momentum modes). These corrections are well studied in related situations and we will be able to adapt existing results to our configurations. 

The $\N=1$ contributions are less well understood. 
We find by direct calculation that string loop corrections due to states localized at image intersections  vanish, for any orbifold and any brane configurations with minimal supersymmetry  in four dimensions (${\cal N}=1$). There is no corresponding nonrenormalization theorem for K\"ahler metrics in minimally supersymmetric field theory. With minimal supersymmetry, there is also no direct argument that string states decouple in the loop, and a priori our expressions contain contributions from massive string states, it is just that the contributions all vanish. Therefore we have really proven a ``string nonrenormalization theorem''. In addition, it is rather rare that statements can be made about minimally supersymmetric orientifolds that hold regardless of orbifold group or brane configuration, but this is one such statement. Another statement of this kind was recently 
made about the  absence of mass renormalization due to $\N=1$ contributions, in \cite{Anastasopoulos:2011kr}.

It would be interesting  to understand how and whether this result can be recast as a statement about symmetries of the theory. Some tentative arguments in this direction can be made, but at the moment we do not know how to cleanly formulate the vanishing result in this paper in terms of symmetries. We will comment on this issue, as well as possible generalizations, in the conclusions. 

In \cite{Berg:2005ja} we pursued a similar calculation neglecting the contributions we consider here. 
There, following the seminal work  \cite{ABFPT}, we first considered a reduction of the relevant (tree-level) effective supergravity using the standard curvilinear coordinates on the covering torus, identified the correct combinations of moduli, then rewrote the worldsheet action in terms of them. By varying this rewritten worldsheet action directly with respect to the moduli fields of interest, we obtained the corresponding vertex operators, that then automatically captured the full moduli dependence. 
In this paper, we found it useful to generalize this strategy somewhat to obtain the moduli dependence for D-branes at angles.  Here, we will not use the standard curvilinear coordinates on the covering tori, but a better choice of coordinate frame is one that is adapted to the D-branes of interest.  A related approach was used to great effect in the work by Hassan \cite{Hassan:2003uq}. Our strategy will essentially be to generalize this to D-branes at angles. Another relevant paper is \cite{Lust:2004}, where similar vertex operators to ours were introduced and used for tree-level string calculations. 

One-loop corrections to gauge couplings have been studied in greater detail than corrections to the K\"ahler metric, and we will make heavy use of \cite{Lust:2003ky} and \cite{Gmeiner:2009fb} for much of the background detail. Although our results will be valid for any orientifold and any brane configuration, we provide some explicit examples in the appendix, focused on the $T^6/\mathbb{Z}_6'$ orientifold and a two-stack configuration (for simplicity). We will not consider any direct phenomenological applications in this paper, but the interested reader may consult \cite{Gmeiner:2009fb} for phenomenologically interesting spectra in the $T^6/\mathbb{Z}_6'$ orientifold with more brane stacks.  

There is plentiful literature on string calculation in toroidal orientifolds, but few make progress on the technical details needed for the kind of calculation we present here. Some examples of papers that develop techniques that are relevant to this work are \cite{Lust:2003ky,Lust:2004,Bertolini:2005qh}.


\section{String effective action}

We begin by an overview of the dimensional reduction of the tree-level effective supergravity action
and how the D-brane moduli appear. Then we review the vertex operators with which one 
can compute this supergravity action as a low-energy limit of string theory. Finally, in sec.\ \ref{n1} and \ref{n2} we use these vertex operators to calculate string amplitudes, from which we extract the one-loop K\"ahler metric of the D-brane moduli. 

\subsection{K\"ahler variables}
\label{Kvariables} 

Let us remind the reader that in effective Type IIB  supergravity the open string moduli are defined as
\be \label{AIIB}
A^\alphai = U_\alphai a^\alphai_1 - a^\alphai_2\ , \quad \mbox{(no sum over $i$)}
\ee
where $U_\alphai$ is the complex structure modulus of the $\alphai$th torus. Under T-duality the complex structure modulus is mapped to the corresponding K\"ahler modulus $T_\alphai$ of the same torus. The form of the open string moduli \eqref{AIIB} can be derived via a dimensional reduction, cf.\ \cite{ABFPT}. In the case of D9/D5-branes the dimensional reduction gives
\be \label{L4}
{\cal L}^{(4)} &=& -\frac12 R^{(4)} + \frac{\partial S \partial \bar S}{(S - \bar S)^2} + \sum_{\alphai=1}^3 \Big[ \frac{\partial U_\alphai \partial \bar U_\alphai}{(U_\alphai - \bar U_\alphai)^2} + \frac{(\partial {\rm Im} T_\alphai)^2}{(T_\alphai - \bar T_\alphai)^2} + \frac{(\partial {\rm Re} T_\alphai +\tfrac12 \sum_{\rm branes} a_1^{\alphai} \stackrel{\leftrightarrow}{\partial} a_2^{\alphai})^2}{(T_\alphai - \bar T_\alphai)^2}\nonumber \\
\mbox{} && + \sum_{\rm branes} \frac{|U_\alphai \partial a^{\alphai}_1 - \partial a^{\alphai}_2|^2}{(T_\alphai - \bar T_\alphai) (U_\alphai - \bar U_\alphai)} \Big]\ .
\ee
Strictly speaking this form of the action would arise in a reduction on a factorized six-torus and some of the complex structure moduli might not be moduli in the orbifold. The form of the open string moduli can be inferred from the last term in \eqref{L4}. It arises from the kinetic term of the gauge fields on the brane, i.e.\ from an expansion of the DBI action. A reduction of the DBI action can be redone in the T-dual type IIA picture with branes at angles. In order to do so, we distinguish the indices of coordinates according to whether they are parallel or tangent to the brane and/or to the non-compact space-time, i.e.\
\be
\label{coordinates}
\mu & : & ||\ \mathbb{R}^{1,3}\ , \nonumber \\
A &:& ||\ {\rm brane\ but} \perp \mathbb{R}^{1,3}\ , \\
a &:& \perp {\rm brane}\ . \nonumber
\ee
The coordinates along the brane are denoted by $\{\xi^\mu, \xi^A\}$ and the coordinates along space-time are $ \{ X^\mu, X^A, X^a \}$. We work in static gauge, i.e.\ $X^A=\xi^A$. To do so, we concentrate on a single representative of an orbit. 

We are interested in expanding $\sqrt{\det (P[G] + {\cal F})}$ to second order in the fluctuations along or transverse to the branes. $P[G] $ stands for the pullback of the metric. 
In appendix A, we obtain for the kinetic term
\beqn
&& \sqrt{\det G_{\mu \nu}}  \sqrt{\det G_{AB}} \sum_{\alphai} \tfrac12 \Big( (\partial_\mu A_\alphai - B_\alphai \partial_\mu \phi^\alphai)(\partial^\mu A_\alphai - B_\alphai \partial^\mu \phi^\alphai) +\rho_\alphai^2 \partial_{\mu} \phi^\alphai \partial^{\mu} \phi^\alphai  \Big) \frac{1}{L^2_{\alphai}}\nonumber \\
&=& \sqrt{\det G_{\mu \nu}}  \sqrt{\det G_{AB}} \sum_{\alphai} \frac{1}{2 L^2_{\alphai}} |T_\alphai \partial \phi^{\alphai} - \partial A_{\alphai}|^2 
\eeqn
with 
\be \label{Tmodulus}
T_\alphai = B_\alphai + i \rho_\alphai\ ,
\ee
where $B_\alphai$ denotes the component of the B-field along the $\alphai$th torus,
$\rho_{\alphai} $ is the volume of the $\alphai$th torus and $L_i$ is the length of the brane along the $\alphai$th torus, as discussed in
more detail in Appendix \ref{app:reduction}.

Analogy to \eqref{AIIB} and \eqref{L4} suggests to define the open string moduli according to 
\be \label{AIIA}
\Phi^\alphai = T_\alphai \phi^\alphai - A_\alphai  \quad \mbox{(no sum over $i$)} \  .
\ee
This is in agreement with formula (3.51) in \cite{Cremades:2003qj}\footnote{Note also the definitions of their $\epsilon$ and $\theta$ on the bottom of page 15 and 17, respectively.}


\subsection{Real vertex operators: no relative angles}

As mentioned in the previous section,
the coordinates $(X^A,X^a)$ are adapted to a specific brane,
with an explicit split into parallel and perpendicular coordinates.
A particular split of this kind will obviously not work for two branes at nonzero angles simultaneously. 
However, for our purposes we only need to insert vertex operators on a single stack of branes.
This is because we are interested in calculating a scalar two-point function and inserting the two vertex operators 
on two different branes in an annulus diagram would lead to a vanishing result after summing over the branes and 
their orientifold images. (We will come back to the different diagrams contributing to the two-point function in sec.\ \ref{setup}.) Therefore, we will only make insertions on a single boundary at a time and our 
variables and vertex operators can be adapted to the stack on which we make the insertion. 

But  correlators of those vertex operators, even if inserted at a single stack, will involve propagators that are determined by boundary conditions at both ends of the open string. So we see that amplitudes will depend on the relative angle between D-branes through the correlators, even though the vertex operators on any given stack may be adapted to that stack.   

One can consider vertex operators for intrinsic D-brane worldvolume fields or for ambient spacetime fields. We will consider spacetime fields, but for completeness we write the relation to 
worldvolume fields in appendix \ref{variables:appendix}, following Hassan \cite{Hassan:2003uq}. 
The vertex operator
for spacetime fields can be read off from  that reference\footnote{To be specific, formula (52) together with (8)-(11) in \cite{Hassan:2003uq}. 
This reference uses a Lorentzian worldsheet,  but we will Wick rotate
to a Euclidean worldsheet.This means $\partial_t \rightarrow i \partial_\tau$. The vertex operators in principle receive an overall $i$ from the integration measure, but this is absorbed in the Euclidean definition of the functional integral.}. We use a plane wave ansatz, e.g.\ $A_M(X)=A_M e^{ip\cdot X}$ for constant $A_M$, to obtain the vertex operators.
Assuming a constant matrix $E_{MN}=G_{MN}+B_{MN}$, the vertex operators
for Wilson lines and D-brane positions are
\beqn \label{Vops}
V_{A_M} & = &{ g_{\rm o} \over \sqrt{2\alpha'}} \int_{\partial \Sigma} d \tau A_M \Big[i  \partial_\tau X^M + \frac{\alpha'}{2} p_N (\psi^N + \eta \tilde{\psi}^N) (\psi^M + \eta \tilde{\psi}^M) \Big] e^{i p \cdot X} \ , \nonumber \\
V_{\phi^M} & = & { g_{\rm o} \over \sqrt{2\alpha'}} \int_{\partial \Sigma} d \tau \phi^M \Big[ i B_{MN} \partial_\tau X^N - G_{MN} \partial_\sigma X^N  \\ && \hspace{2cm} + \frac{\alpha'}{2} p_K (\psi^K + \eta \tilde{\psi}^K) (E_{MN} \psi^N - \eta E^T_{MN} \tilde{\psi}^N) \Big] e^{i p \cdot X} \ , \nonumber 
\eeqn
where $g_{\rm o}$ is the (dimensionless) open string coupling, and
the $\alpha'$ factor is such that the vertex operators are dimensionless (for the normalization factor see \cite{Polchinski:1998rq}).
Also $\eta=\pm 1$  is defined to take the same value at both ends in the Ramond sector and the opposite value at the two ends in the Neveu-Schwarz sector. Without loss of generality, we assign\footnote{This is the same choice as in Polchinski \cite{Polchinski:1998rr}, Ch. 10,
which he calls $\nu'=0$.}
\beqn \label{eta}
\eta = \left\{ \begin{array}{ccl}
+1 & , & \sigma = \pi \\
\left\{ \begin{array}{cc}
+1 & ({\rm R}) \\
-1  & ({\rm NS})
\end{array}
\right\}
& , & \sigma = 0 \ .
\end{array} \right. 
\eeqn
We emphasize  that the sign combinations in the above vertex operators are defined to be T-duality covariant. For example, using the T-dual coordinate  $X'(w,\bar{w})= X_L(w)-X_R(\bar{w})$ one sees that $\partial_{\sigma} \rightarrow -i\partial_{\tau}$,
which enforces the above sign relation between $\partial_{\tau}X$ in $V_{A_M}$ and $\partial_{\sigma}X$ in $V_{\phi_M}$. 
For vanishing $B$-field background (which we assume from now on), the vertex operators for the Wilson lines
are the same as above, but the 
position scalars simplify to
\beqn \label{Vphi}
V_{\phi_N} & = & - { g_{\rm o} \over \sqrt{2\alpha'}}\int_{\partial \Sigma} d \tau \phi_{N} \Big[ i \partial_\sigma X^N - \frac{\alpha'}{2} p_K (\psi^K + \eta \tilde{\psi}^K) (\psi^N - \eta \tilde{\psi}^N) \Big] e^{i p \cdot X}\ , 
\eeqn
where
\beqn \label{phisubN}
\phi_N & = & \phi^M G_{MN} \ .
\eeqn
In static gauge, the only non-vanishing components of the fields with lower indices are $\phi_a$ and $A_A$, cf.\ eq.\ \eqref{coordinates} for the notation, and also appendix \ref{variables:appendix}. (That is, $\phi_A=A_{a}=0$, but we emphasize that this is a gauge-dependent statement, as explained in detail
in \cite{Hassan:2003uq}.) Moreover, we only consider momentum along the non-compact directions so that the only non-vanishing momentum components are $p_\mu$. Thus, the vertex operators become
\beqn \label{Vstatic0}
V_{A_A} & = & { g_{\rm o} \over \sqrt{2\alpha'}}\int_{\partial \Sigma} d \tau A_A \Big[ i \partial_\tau X^A + \frac{\alpha'}{2} p_\mu (\psi^\mu + \eta \tilde{\psi}^\mu) (\psi^A + \eta \tilde{\psi}^A) \Big] e^{i p \cdot X} \ , \nonumber \\
V_{\phi_a} & = & -{ g_{\rm o} \over \sqrt{2\alpha'}} \int_{\partial \Sigma} d \tau \phi_{a} \Big[  \partial_\sigma X^a - \frac{\alpha'}{2} p_\mu (\psi^\mu + \eta \tilde{\psi}^\mu) (\psi^a - \eta \tilde{\psi}^a) \Big] e^{i p \cdot X}\ .
\eeqn
These can be rewritten in a more familiar form using the boundary conditions. To do so, we recall that the relations between the left- and right-moving fermions in Neumann and Dirichlet directions, respectively, are
\beqn
\left. \psi^{\mu} \right|_{\sigma = 0, \pi} &=& \eta\, \left. \tilde{\psi}^{\mu}\right|_{\sigma = 0, \pi} \hspace{1cm} (\mbox{Neumann)}\ , \nonumber \\
\left. \psi^A \right|_{\sigma = 0, \pi} &=& \eta\, \left. \tilde{\psi}^A\right|_{\sigma = 0, \pi} \hspace{1cm} ({\rm Neumann})\ , \nonumber \\
\left. \psi^a\right|_{\sigma = 0, \pi} &=& - \eta\, \left. \tilde{\psi}^a\right|_{\sigma = 0, \pi} \hspace{.7cm} ({\rm Dirichlet})\ ,
\eeqn
where of course both sides have to be taken at the same value of $\sigma$, i.e.\ both at $\sigma = 0$ or both at $\sigma = \pi$. 
Using this we find
\beqn \label{Vstatic}
V_{A_A} & = & { g_{\rm o} \over \sqrt{2\alpha'}} \int_{\partial \Sigma} d \tau A_A \Big[ i \partial_\tau X^A +2 \alpha' (p \cdot \psi) \psi^A  \Big] e^{i p \cdot X} \ , \nonumber \\
V_{\phi_a} & = & -{ g_{\rm o} \over \sqrt{2\alpha'}} \int_{\partial \Sigma} d \tau \phi_{a} \Big[  \partial_\sigma X^a - 2\alpha' (p \cdot \psi) \psi^a  \Big] e^{i p \cdot X}\ .
\eeqn
The 
 reason that each pair of terms in the fermions added up instead of cancelling (which would have been the other possibility) is that the combinations of worldsheet fields in the vertex operators are {\it defined} as the pieces
that are nonvanishing under the given boundary conditions (see appendix \ref{variables:appendix}). 
However, since we have now made explicit use of the boundary condition
for $\psi$, the distinction is no longer manifest. 

 To keep this simple holomorphic form of the vertex operators but
 still impose boundary conditions in a way that makes no explicit reference
 to right-movers
  we use the well-known
 ``doubling trick" (see e.g.\ \cite{Polchinski:1998rq}).
The trick is to define a ``doubled'' holomorphic fermion field extending 
into the ``unphysical'' region $\pi < \sigma \le 2 \pi$ by using the right-mover $\tilde{\psi}$ there:
\beqn
\psi^A(\sigma, \tau) &=& \left\{ \begin{array}{ccc}
\psi^A(\sigma, \tau) & , & 0 \le \sigma \le \pi \\
\tilde{\psi}^A(2 \pi - \sigma, \tau) & , & \pi \le \sigma \le 2 \pi\ ,
\end{array} 
\right. \nonumber \\
\psi^a(\sigma, \tau) &=& \left\{ \begin{array}{ccc}
\psi^a(\sigma, \tau) & , & 0 \le \sigma \le \pi \\
-\tilde{\psi}^a(2 \pi - \sigma, \tau) & , & \pi \le \sigma \le 2 \pi\ .
\end{array} 
\right. 
\eeqn
The boundary condition at $\sigma = \pi$ is automatically fulfilled, while the condition at $\sigma = 0$ amounts to the periodicity conditions
\beqn
\psi^A(2 \pi, \tau) &=& \eta \psi^A(0, \tau) \ , \nonumber \\
\psi^a(2 \pi, \tau) &=& \eta \psi^a(0, \tau) \ .
\eeqn
Then we can use the same expressions for the vertex operators 
as for the physical left-movers above, but now 
where $\psi^a$ and $\psi^A$ are holomorphic and defined over the full range 0 to $2\pi$. 
%

\subsection{Complex vertex operators: angles}

In this section we write down vertex operators directly for the 
complex variables $\Phi^i$ defined in \eqref{AIIA}. It might be worthwhile to contrast our approach with that of \cite{Bertolini:2005qh}. Those authors consider vertex operators with branch cuts in the complex plane. We work entirely in cylinder variables, where there is no branch cut. Instead there is quasiperiodicity as a vertex operator crosses through the unphysical region --- in the sense of the method of images --- back into the physical region (see figure \ref{unphysical}) below. Another difference is that those authors begin by performing real matrix rotations, and then diagonalize to obtain complex embedding coordinates. They can reproduce the DBI action as output. We obtain our variables by comparing with the reduction of the DBI action as input, but we consider adapted coordinates, so we do not need to diagonalize. Ultimately, whatever approach one prefers they  should  be equivalent, and 
indeed we reproduce their results on D-branes at angles,
for example \eqref{eq:Uangle} (with the replacement $U\rightarrow -1/U$, as we T-dualize on a different axis). For other work in this direction see also \cite{Epple:2004nh}.

\begin{figure}
\begin{center}
\includegraphics[width=0.5\textwidth]{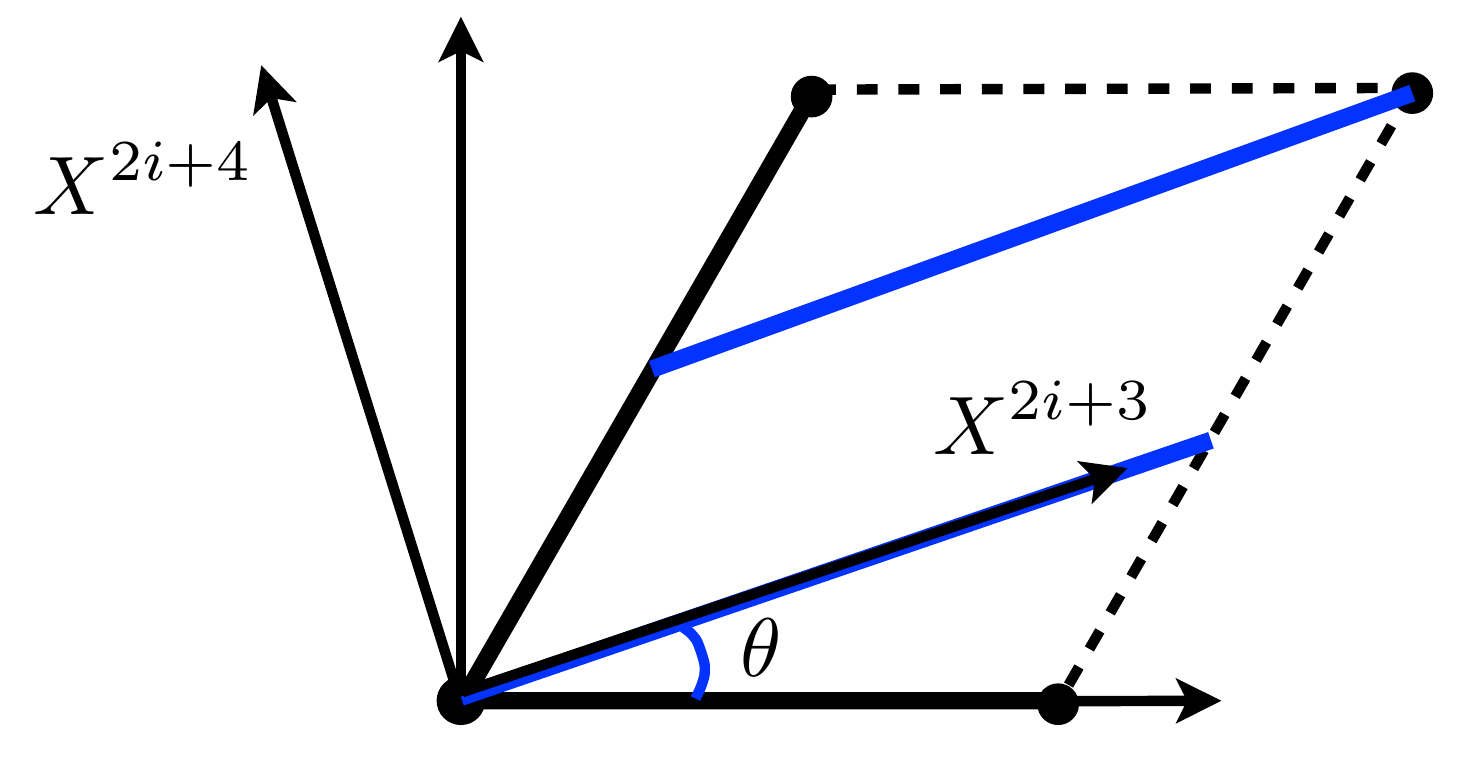}
\caption{Tilted coordinates for a brane with wrapping number $(n,m)=(2,1)$. The complex coordinates $Z_{\theta}^i$ and $\Psi_{\theta}^i$ also include the length $L_i$ and perpendicular distance $D_i$. }
\label{Xcoords}
\end{center}
\end{figure}

Now on to the calculation. We have to introduce complex coordinates 
along the internal tori. 
In the conventions of \cite{Polchinski:1998rq}, we have $
\partial_{\sigma} = \partial+\bar{\partial}$ and
$\partial_{\tau} = i(\partial-\bar{\partial}) $, using which the boundary conditions are
\be \label{BC}
(\partial+\bar{\partial}) X^A &=& 0  \qquad \mbox{Neumann, along brane} \\
(\partial-\bar{\partial})X^a &=& 0  \qquad \mbox{Dirichlet, perpendicular to brane. } 
\ee
Now for our variables
\be
Z_{\theta}^i  &=& \frac{1}{\sqrt{2}} \Big(L_i X^{2i+3} + i D_i X^{2i+4}\Big)\ , \label{Zi} \\
\Phi_i &=&  T_i \phi^i - A_i  \label{Phi}   \quad \mbox{(no sum over $i$)} \ , 
\ee
where $X^{2i+3}$ is a coordinate along the $i$th torus parallel to the brane (stack) and $X^{2i+4}$ is transverse to it, i.e.\ Ê$X^{2i+3}$ is one of the coordinates $X^A$ and $X^{2i+4}$ is one of the coordinates $X^a$, but the present notation emphasizes the relation to the $i$th torus, cf.\ fig.\ \ref{Xcoords}. 
In \eqref{Zi}, $L_i$ is the length of the brane along the $i$th torus and $D_i$ is the distance to the neighboring parts of the brane along the $i$th torus. For more detail,
see fig.\  \ref{tiltedtorus} and eqs.\ \eqref{LD} in the appendix. The normalization in \eqref{Zi} was chosen in order to ensure 
\be
\langle \partial Z_{\theta}^i(z) \partial \bar Z_{\theta}^{\bar{\jmath}}(w) \rangle = \frac{\delta^{ij}}{|z-w|^2}
\ee
at disk level, when normalizing the coordinates in such a way that 
\be
\langle \partial X^I(z)  \partial X^J(w) \rangle =  \frac{G^{IJ}}{|z-w|^2}\ .
\ee
Here, $X^I,X^J$ stand for any of the internal coordinates and the internal metric is given by a product of three factors of the diagonal form given in \eqref{t2metric3}. 

Using \eqref{Zi}, the boundary conditions \eqref{BC} can be rephrased as 
\be \label{BCcomplex}
\bar{\partial}\bar{Z_{\theta}}^{\bar{\imath}} &=& -\partial Z_{\theta}^i\ , \\
\bar{\partial} Z_{\theta}^i &=& - \partial \bar{Z_{\theta}}^{\bar{\imath}} \ . 
\ee

Inverting the expressions \eqref{Zi} and \eqref{Phi} and using $T=i T_2$ for a background without $B$-field gives
\be \label{Aphi}
A_i  &=& -\frac12 (\Phi_i + \bar{\Phi}_{\bar{\imath}})\ , \nonumber \\
\phi^i &=& {\Phi_i - \bar{\Phi}_{\bar{\imath}} \over 2 i (T_i)_2} 
\ee
and
\be \label{x1x2}
X^{2i+3} &=& \frac{1}{\sqrt{2} L_i} (Z_{\theta}^i+\bar{Z_{\theta}}^{\bar{\imath}})\ , \\
X^{2i+4} &=& \frac{1}{\sqrt{2} i D_i} (Z_{\theta}^i-\bar{Z_{\theta}}^{\bar{\imath}}) \; .
\ee
Note that the boundary action and the corresponding vertex operators \eqref{Vstatic} involve the position variables $\phi$ with a lower index instead of an upper index. Using the metric \eqref{t2metric3}, this can be obtained as
\be
\phi_i = D_i^2 {\Phi_i - \bar{\Phi}_{\bar{\imath}} \over 2 i (T_i)_2}\ .
\ee

Now we have
\be \label{Lbdryboson} 
{\mathcal L}_{\rm bdry}^X &\sim& iA_A \partial_{\tau}X^A - \phi_a \partial_{\sigma}X^a = iA_i \partial_{\tau}X^{2i+3} - \phi_i \partial_{\sigma}X^{2i+4} \\
&=& -\frac{i}{2} (\Phi_i + \bar{\Phi}_{\bar{\imath}}) i(\partial-\bar{\partial}) \frac{1}{\sqrt{2} L_i} (Z_{\theta}^i+\bar{Z}_{\theta}^{\bar{\imath}}) 
- D_i^2 {\Phi_i - \bar{\Phi}_{\bar{\imath}} \over 2 i (T_i)_2} (\partial+\bar{\partial}) \frac{1}{\sqrt{2} i D_i} (Z_{\theta}^i-\bar{Z}_{\theta}^{\bar{\imath}}) \\
&=& \frac{1}{2 \sqrt{2} L_i} \left[ (\Phi_i + \bar{\Phi}_{\bar{\imath}}) (\partial-\bar{\partial}) (Z_{\theta}^i+\bar{Z}_{\theta}^{\bar{\imath}}) + (\Phi_i - \bar{\Phi}_{\bar{\imath}}) (\partial+\bar{\partial}) (Z_{\theta}^i-\bar{Z}_{\theta}^{\bar{\imath}}) \right] \\
&=& \frac{1}{\sqrt{2} L_i} \left[ \Phi_i (\partial Z_{\theta}^i - \bar \partial \bar{Z}_{\theta}^{\bar{\imath}}) + \bar{\Phi}_{\bar{\imath}} (\partial \bar{Z}_{\theta}^{\bar{\imath}} - \bar \partial Z_{\theta}^i) \right]\ ,
\ee
where in the third line we used $D_i L_i = (T_i)_2$, cf.\ \eqref{LD}.
Using \eqref{BCcomplex}, this can be rewritten as
\be \label{LbdryX}
{\mathcal L}_{\rm bdry}^X &\sim& \frac{\sqrt{2}}{L_i} \left[\Phi_i \partial Z_{\theta}^i + \bar{\Phi}_{\bar{\imath}} \partial \bar{Z}_{\theta}^{\bar{\imath}} \right]\ .
\ee

Before considering the fermions, let us mention that we of course did not have to use  coordinates 
$Z_{\theta}$ adapted to a specific brane.  Another obvious choice would be fixed orthogonal coordinates corresponding to basis vectors along the horizontal and vertical axis in each internal plane. Then $\theta$ would correspond to the rotation angle of the brane with respect to the horizontal axis, cf.\ fig.\ \ref{Xcoords},
and the relation to those coordinates $Z$ is simply
\be \label{Zrot}
Z_{\theta} = e^{-i\theta}Z \; .
\ee
\begin{figure}
\begin{center}
\includegraphics[width=0.5\textwidth]{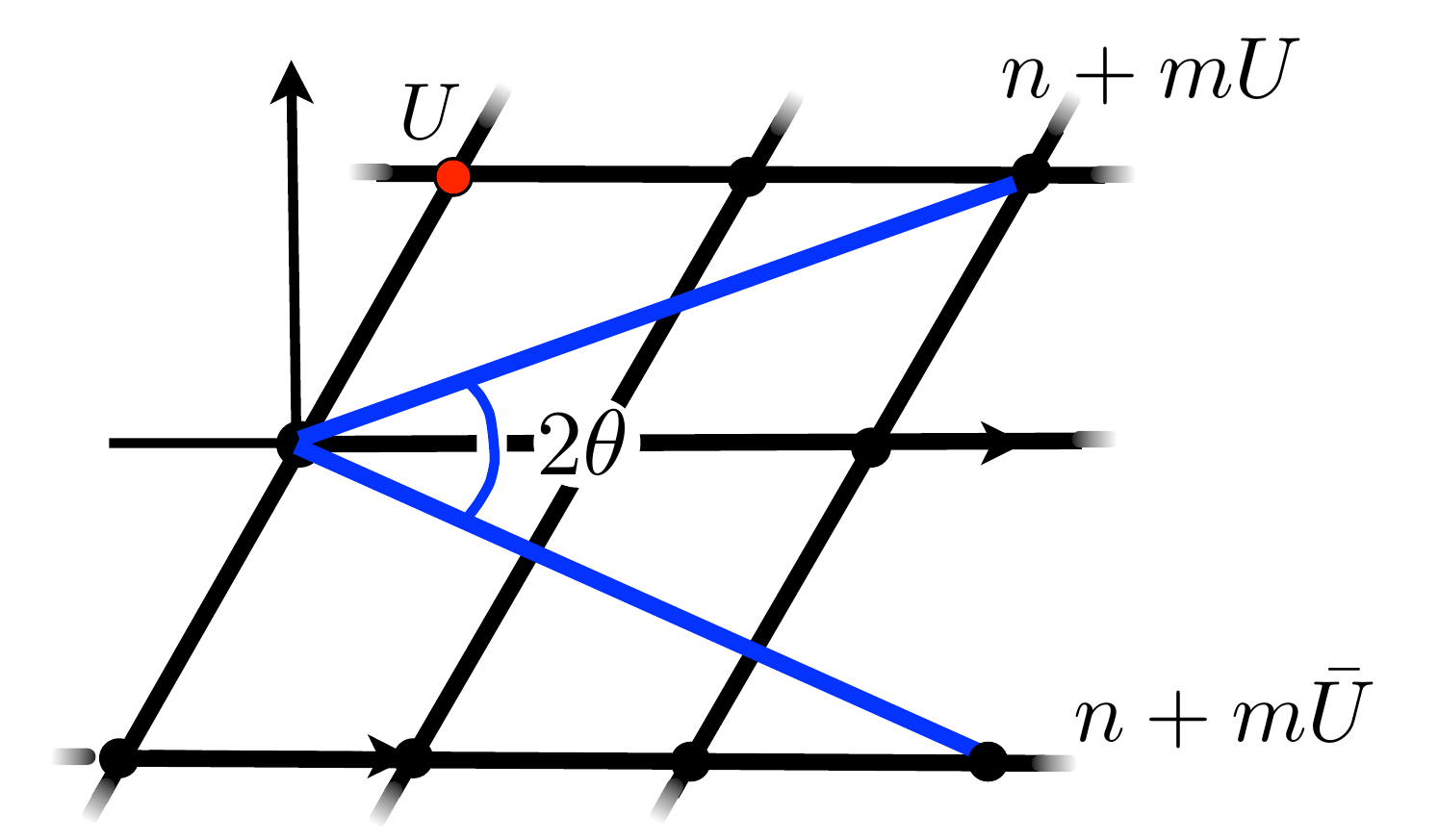}
\caption{Covering space. Since $|n+mU|=|n+m\bar{U}|$,
dividing $n+mU$ by $n+m\bar{U}$ gives just the angle $e^{2i\theta}$.
In this example $(n,m)=(2,1)$. }
\label{Uangle}
\end{center}
\end{figure}
It is of course for this reason that we put the subscript $\theta$ on our adapted $Z_{\theta}$ coordinate. In the 
un-adapted $Z$ coordinate, the boundary conditions become
\be
\bar{\partial}\bar{Z} &=& -e^{-2i\theta}\partial Z\ , \label{BCtransform} \\
\bar{\partial} Z &=& -e^{2i\theta} {\partial} \bar{Z}\  \label{BCtransform2} . 
\ee
In the orbifold the angle is fixed by the wrapping numbers and the complex structure of the spacetime torus. We see 
in figure \ref{Uangle} that
\be  \label{eq:Uangle}
 e^{i   \theta}=\sqrt{\frac{n+mU }{n+m\bar U }}  \ .
\ee
For now we will keep the adapted coordinate $Z_\theta$ defined in \eqref{Zi}.  

We use the same logic to obtain the fermion contribution to the vertex operator as we did for the bosonic part. We define
\be
\Psi_{\theta}^i = \frac{1}{\sqrt{2}} \Big(L_i \psi^{2i+3} + i D_i \psi^{2i+4}\Big) \quad , \quad \tilde \Psi_{\theta}^i = \frac{1}{\sqrt{2}} \Big(L_i \tilde \psi^{2i+3} + i D_i \tilde \psi^{2i+4} \Big)\ .
\ee
This leads to 
\be
\psi^{2i+3} &=& \frac{1}{\sqrt{2} L_i} (\Psi_{\theta}^i+\bar{\Psi}_{\theta}^{\bar{\imath}})\ , \label{psiPsi1}\\
\psi^{2i+4} &=& \frac{1}{\sqrt{2} i D_i} (\Psi_{\theta}^i-\bar{\Psi}_{\theta}^{\bar{\imath}}) \label{psiPsi2}
\ee
and the same for the quantities with tildes. Using this and \eqref{Aphi} and 
we obtain for the fermionic contribution (suppressing a factor $\alpha'p_{\mu}\psi^{\mu}$)
\be
{\mathcal L}_{\rm bdry}^{\psi} &\sim& A_A (\psi^A + \eta \tilde \psi^A) + \phi_a (\psi^a - \eta \tilde \psi^a) \\
&=& - \frac{1}{\sqrt{2} L_i} \left[ \Phi_i (\Psi_{\theta}^i +\eta \bar{\tilde{\Psi}}_{\theta}^{\bar{\imath}}) + \bar{\Phi}_{\bar{\imath}} (\bar{\Psi}_{\theta}^{\bar{\imath}} +\eta \tilde{\Psi}_{\theta}^i) \right]\ .  \label{Lbdryfermion}
\ee
This can still be simplified by using the boundary conditions
\be
\psi^A = \eta \tilde \psi^A \quad , \quad \psi^a = - \eta \tilde \psi^a\ ,
\ee
which can be rewritten using \eqref{psiPsi1} and \eqref{psiPsi2} as
\be \label{cplxBC}
 \bar \Psi_{\theta}^{\bar{\imath}} = \eta \tilde{\Psi}_{\theta}^i \ .
\ee
Using this in \eqref{Lbdryfermion}, we finally obtain
\be
{\mathcal L}_{\rm bdry}^{\psi} &\sim& A_A (\psi^A + \eta \tilde \psi^A) + \phi_a (\psi^a - \eta \tilde \psi^a) \\
&=& - \frac{\sqrt{2}}{L_i} \left[ \Phi_i \Psi_{\theta}^i + \bar{\Phi}_{\bar{\imath}} \bar{\Psi}_{\theta}^{\bar{\imath}} \right]\ .\nonumber
\ee
We now have the result for bosons and fermions,
and reinstating the factor $\alpha'p_{\mu}\psi^{\mu}$ the total vertex operators are (no summation over $i$)
\be \label{Vtotal}
V_{\Phi_i} =\frac{g_o}{\sqrt{\alpha'} L_i} e_i \left[ \partial Z_{\theta}^i - \alpha'p_{\mu}\psi^{\mu}\ \! \Psi_{\theta}^i \right]e^{ip\cdot X}\quad , \quad V_{\bar{\Phi}_{\bar{\imath}}} =\frac{g_o}{\sqrt{\alpha'} L_i} \bar{e}_{\bar{\imath}} \left[ \partial \bar{Z}_{\theta}^{\bar{\imath}} - \alpha' p_{\mu} \psi^{\mu}\ \! \bar{\Psi}_{\theta}^{\bar{\imath}} \right]e^{ip\cdot X}\ ,
\ee
where $e_i$ denotes the polarisation in field space. We emphasize again that these vertex operators make use of the coordinates which are adapted to a particular brane (stack). Only in the special case that all vertex operators are inserted on the same brane, as we will consider in the rest of this paper, can they be directly applied. This does not work for non-planar amplitudes, where vertex operators are inserted on different branes at relative angles (which might be image branes under the orientifold action). In those cases one would have to rotate the coordinates $Z_{\theta}$ (and similarly the fermions $\Psi_{\theta}$) like in \eqref{Zrot} with an angle appropriate for the brane on which the vertex operator is inserted. We note that even for planar amplitudes where \eqref{Vtotal} can be used, the {\it correlators} will always depend explicitly on the relative angle, as we will show momentarily. 

Before proceeding further, however, let us perform a quick test of our vertex operators \eqref{Vtotal}, by reproducing the known moduli dependence at tree level. At disk level the correlator has the moduli dependence
\be
\langle V_{\Phi_i} V_{\bar{\Phi}_{\bar{\imath}}} \rangle \sim \frac{e^{-\Phi_{10}}}{L_i^2} \prod_j L_j \ ,
\ee
where the factor $e^{-\Phi_{10}}$ comes from the usual dilaton dependence of a disk amplitude and $\prod_j L_j$ is the volume of the cycle wrapped by the brane under consideration. This factor arises from the integration over the zero modes\footnote{The integral over zero modes $x_0^\mu$ usually produces a delta function in spacetime momenta, cf.\ (6.2.13) in \cite{Polchinski:1998rq}. However, in our case there is no momentum along the $X^A$ directions, so the zero modes $x^A_0$ drop out of the integrand. Then, the integral simply gives the volume of the three-cycle that the brane wraps.} of $X^A$. In order to obtain the moduli dependence of the K\"ahler metric, one should perform a Weyl rescaling, leading to an additional factor of $e^{2 \Phi_{4}}=e^{2 \Phi_{10}} {\rm Vol}^{-1}$, where ${\rm Vol}=\prod_j (T_2)_j$ is the volume of the Calabi-Yau orientifold. This results in
\be
G_{\Phi_i \bar{\Phi}_{\bar{\imath}}}^{{\rm disk}} \; \sim \; \frac{e^{\Phi_{10}}}{L_i^2}  \prod_{j=1}^3 \frac{L_j}{(T_2)_j} = \frac{e^{\Phi_{4}}}{L_i^2} \prod_{j=1}^3 \frac{L_j}{\sqrt{(T_2)_j}} = \frac{e^{\Phi_{4}}}{V_i (T_2)_i} \prod_{j=1}^3 \sqrt{V_j}\ ,
\ee
where we used \eqref{LD} (together with \eqref{Tdef}) and \eqref{Vaa}. This moduli dependence precisely agrees with known results,  for example eq.\ (53) in \cite{Honecker:2011sm}.

We now go through the arguments of the doubling trick again for the complexified fermion.
As in \eqref{cplxBC}, the boundary condition at each end can have a phase associated with that end,
and we now emphasize this by an index on the angle $\theta$: 
\be
 \overline{\Psi}_{\theta_0}  &=& \eta\,  {\tilde{\Psi}}_{\theta_0}\hspace{1cm} \mbox{(at $\sigma=0$)}\ , \label{rot1} \\
\overline{\Psi}_{\theta_{\pi}} &=& \eta\,  {\tilde{\Psi}}_{\theta_{\pi}}\hspace{1cm} \mbox{(at $\sigma=\pi$)} \ .
\ee
We now want to express them in terms of a single field, let us say adapted to the brane at angle $\theta_0$. To do so we simply rotate
$\Psi_{\theta_{\pi}}=e^{-i(\theta_{\pi}-\theta_{0})}\Psi_{\theta_0}$, $\tilde{\Psi}_{\theta_{\pi}}=e^{-i(\theta_\pi-\theta_{0})}\tilde{\Psi}_{\theta_0}$ to obtain
\be
 \overline{\Psi}_{\theta_0}  &=& \eta\, {\tilde{\Psi}}_{\theta_0}\hspace{1cm} \mbox{(at $\sigma=0$)}\ , \label{rot2} \\
\overline{\Psi}_{\theta_{0}} &=& \eta\, e^{- 2i(\theta_{\pi}-\theta_{0})}  \tilde{\Psi}_{\theta_0} \hspace{1cm} \mbox{(at $\sigma=\pi$)}\ .
\ee
 The doubling trick
 again extends the fermion into the ``unphysical'' region $\pi < \sigma \le 2 \pi$ by using 
 a translated right-mover $\tilde{\Psi}_{\theta}$  there:%
\be  \label{doubling}
\Psi_{\theta_0}(w) &=& \left\{ \begin{array}{ccc}
\Psi_{\theta_0}(w) & , & 0 \le {\rm Re} \, w  \le \pi \\
\eta e^{ 2i(\theta_{\pi}-\theta_{0})}\overline{\tilde{\Psi}}_{\theta_0}(2 \pi -\bar{w}) & , & \pi \le {\rm Re} \, w \le 2 \pi\ .
\end{array} 
\right. 
\ee
The boundary condition at ${\rm Re} \, w = \pi$ is now fulfilled by construction, while the condition at ${\rm Re} \, w = 2\pi$ 
becomes $\Psi_{\theta_0}(2 \pi) =\eta e^{ 2i(\theta_{\pi}-\theta_{0})} \overline{\tilde{\Psi}}_{\theta_0}(0)$, which 
using \eqref{rot1} turns into the quasiperiodicity condition
\be  \label{quasi}
\Psi_{\theta_{0}}(2 \pi) =  e^{2 i (\theta_{\pi}-\theta_{0})} \Psi_{\theta_{0}}(0) \ .
\ee
So this is a condition on the doubled holomorphic field. Therefore
it is best interpreted on the  covering torus. 
The doubling trick for the complex bosons $Z$ works completely analogously. 

To summarize, the angle difference $2(\theta_{\pi}-\theta_{0})$ appears as a twist of the boundary condition
in the horizontal direction on the covering torus. See figure \ref{unphysical} for an illustration. 
This means that even if the direct dependence on the angle of the brane can be rotated away 
for vertex operators inserted at only a single boundary (using $Z$ and $\Psi$ above),
the {\it relative} angle of rotation between two branes will still appear in 
{\it correlators} of the holomorphic complex fields $Z$ and $\Psi$, since they must display the requisite quasiperiodicity
\eqref{quasi}. This is of course completely obvious physically; only the relative angle between branes
should ultimately affect physical results. 

It may be useful to note that if one insists on working with a physical 
 fundamental region, without the doubling trick, there is a formal 
asymmetry between  worldsheet bosons and worldsheet fermions. In particular, the angle
appears  in correlators of bosons, but  for the fermions,
the angle is instead hidden in the boundary relation between $\Psi$ and $\tilde{\Psi}$.\footnote{Compare e.g. 
eq.\ (B.9) in the appendix of \cite{Bertolini:2005qh}. There,
only the correlator of bosons depends on the open string metric, which in our T-duality frame means
it depends on the angle. The correlator of holomorphic fermions,
on the other hand, depends only on the closed string metric,
i.e.\ without the angle.}
With the doubling trick, the angle appears in correlators of fermions and bosons in the same way.
We emphasize that this is only a matter of convenience and either point
of view may be adopted. 

\subsection{One-loop effective action}

\begin{figure}
\begin{center}\includegraphics[width=0.7\textwidth]{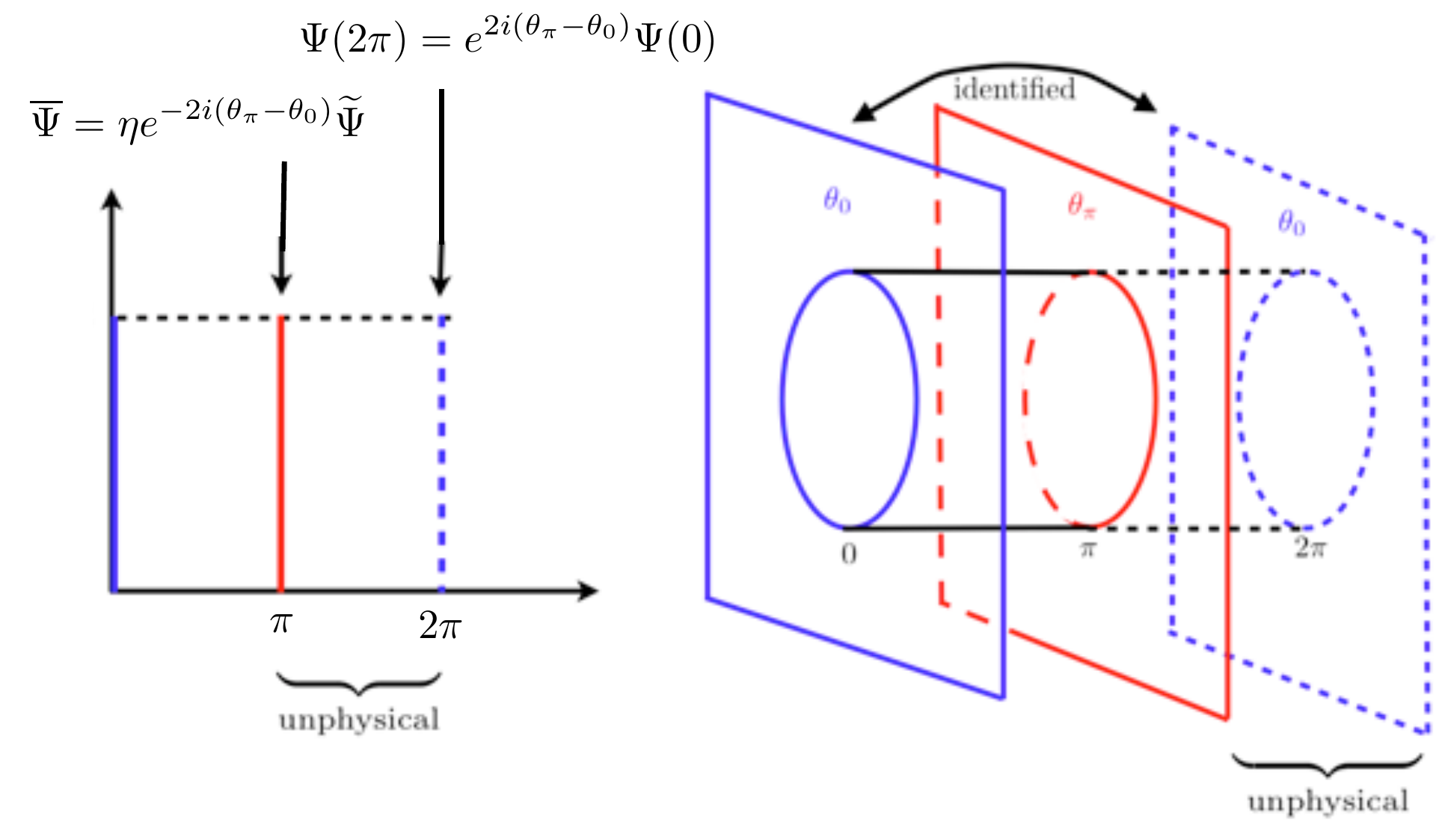}
\vspace{-3mm}
\caption{As the physical cylinder worldsheet only extends between $0$ and $\pi$, the quasiperiodicity of 
our extended fields on the covering torus
lies in the ``unphysical'' region (in the sense of the method of images).
Notice that the branes at $\sigma=0$ and $\sigma=\pi$ may 
be at angles, (i.e.\ $\theta_0\neq\theta_{\pi}$) but this is not drawn in the figure.}
\label{unphysical}
\end{center}
\end{figure}
We now discuss what contributions to the
one-loop effective action are expected to be nonzero.
First of all, we do expect moduli-dependent $\N=2$ contributions, as discussed in the introductions
by combining the arguments of \cite{Berg:2005ja} and \cite{Lust:2003ky}. With
angles we may also expect additional $\N=1$ contributions
that had no analogy without angles. 

The $\N=2$ contributions arise if some of the branes and orientifold planes are parallel to each other 
along one of the three tori. Thus, the corrections are very similar to the case without angles, as will be discussed in more detail in sec.\ \ref{n2}. Focusing on the dependence on the K\"ahler moduli $T$, from \cite{Berg:2005ja}
we expect a K\"ahler potential correction that has a term quadratic in the D-brane scalar $\Phi$:
\be
\Delta K_{g_{\rm s}}(\Phi,\bar{\Phi},T,\bar{T}) \sim f(T) \Phi\bar{\Phi} + \ldots
\ee
for some function $f(T)$. From eq.\ (2.77) of \cite{Berg:2005ja} we may guess
\be
\Delta K_{g_{\rm s}}(\Phi,\bar{\Phi},T,\bar{T}) \sim 
 -E_2''(0,T)\Phi\bar{\Phi} + \ldots
\ee
where $E_2$ is the generalized
nonholomorphic Eisenstein series and the derivative is with respect to the first argument $\Phi$.
Using the identity\footnote{This is eq.\  (C.32) and (C.15) in \cite{Berg:2005ja}, and we also use
(C.17) in \cite{Berg:2005ja}, but note that the latter has a spurious $1/T_2$.} $E_2''(0,T)=-{2\pi^2 i \over T-\bar{T}} \tilde{E}_1(0,T)$ we see that in fact
\be \label{expect}
f(T) \sim   \frac{\tilde{E}_1(0,T)}{T_2} \sim \frac{1}{T_2} \Big( \ln (|\eta(T)|^4 T_2)  + \mbox{($T$-independent terms)} \Big)\ ,
\ee
where $\eta$ is the Dedekind eta function. (This can also be understood
somewhat more indirectly through well-known moduli-dependent corrections to the gauge coupling,
by using ${\N}=2$ supersymmetry.) 
We will see the form \eqref{expect} in section \ref{n2}.  To discuss the ${\N}=1$ contributions,
we need to introduce more detail. 


\section{Setup}
\label{setup}

We are considering an arbitrary type IIA $T^6/\mathbb{Z}_N$-orientifold with D6-branes at angles,
but we do not see any reason that our results
would not generalize immediately to e.g.\ $T^6/(\mathbb{Z}_N\times \mathbb{Z}_M) $. The 
orientifold group is generated by the orbifold generator $\twist$ and $\Omega \cR$, where $\Omega$ is the 
worldsheet parity operator and $\cR$ corresponds to a reflection along the $x$-axes of the three tori. The background 
contains brane stacks $[a]$ and orientifold planes $O_{k}$. Here, $[a]$ stands for the whole orientifold orbit, i.e.\ 
together with a particular brane $a$, it also contains all images under the orientifold group actions, 
i.e.\ $[a] = \{ a_k = \twist^k a, \cR a_k; k=0,\ldots , N-1\}$. Moreover, the orientifold planes $O_{k}$ 
lie along the three dimensional submanifolds which are kept fixed by the action of $\cR \twist^k$. This implies that
the image $\cR a_k$ can be obtained from $a$ by a reflection along $O_{k}$.
In this paper we only consider so-called {\it bulk} branes, i.e.\
orientifold invariant combinations of D-branes. 
This is to be contrasted with {\it fractional} branes, that are themselves localized at fixed points. 
More detail in the specific example of $T^6/{\mathbb Z}_6'$ is  provided in appendix \ref{app:tadpole}.

In this section we would like to review which amplitudes contribute to the 2-point function of 
the open string positions $\phi_a$ and Wilson-line scalars $A_A$. The presentation closely follows 
the analogous discussion for gauge-coupling corrections in sec.\ 2.2 of \cite{Lust:2003ky}.
The first point to note is that since we are interested in the 2-point 
function of certain open string scalars, only worldsheets with a boundary can contribute, i.e.\ the 
annulus and M\"obius diagrams. 

For the annulus amplitude, only strings with no 
insertion of the orbifold operator $\twist^k$ can contribute. (We note that this
is quite different from configurations with no angles, when all twists can in principle contribute.)
The argument is the following.
As is well known \cite{Angelantonj:2002ct,Polchinski:1998rq} 
the trace in the open string channel is written
\be \label{Ann}
\sum_k \sum_{a,b} \langle a,b | q^{H} \twist^k | a,b \rangle\ ,
\ee
where $a$ and $b$ stand for the branes on which the open string state starts and ends, respectively. 
More precisely, $b$ could either be a brane in the same orientifold orbit as $a$ or in a different orbit. 
Now, $\twist^k | a,b \rangle = | \twist^ka,\twist^kb \rangle$, and this is
really a different open string state than $|a,b\rangle$. So the rotated
strings do not contribute to the trace. Strictly speaking this argument does
not hold for $k=\frac{N}{2}$ (for even $N$) in which case $\twist^{N/2} | a,b \rangle = | a, b \rangle$.
However, the $k=\frac{N}{2}$-sector contribution vanishes due to cancellation of twisted tadpoles, which imposes
\be
{\rm Tr}\, \gamma^a_{\twist^{N/2}} = 0
\ee
on the Chan-Paton factors \cite{Blumenhagen:1999ev}. Thus, only the untwisted annulus 
amplitudes with $k=0$ can contribute in principle. 

Among these, one has to distinguish amplitudes for which both open string ends lie on branes with 
non-vanishing relative angle along all three tori, leading to $\N=1$ contributions, and those for which the 
two branes are parallel along one of the three tori, leading to the $\N=2$ contributions discussed in the 
last section. We will see in section \ref{n1} that the $\N=1$ contributions actually vanish and we come 
back to the $\N=2$ contributions in section \ref{n2}. 

We now make a general comment. In \eqref{Ann} we could move the finite 
sum over $k$ into the trace to exhibit the projector
\be
P_{\rm orbifold} = \sum_{k=0}^{N-1} \Theta^k |a,b\rangle \; . 
\ee
Because it is a projector (i.e.\ $P_{\rm orbifold}^2=P_{\rm orbifold}$), 
 the annulus amplitude
 \eqref{Ann}  then only propagates
 invariant (orbifold-neutral) states $P_{\rm orbifold}|a,b\rangle$, in the open string channel.
 But  in actual calculation, the trace is performed without moving the sum over $k$ into the trace. That is, it is calculated for each orbifold-charged
 sector of the theory running in the open-string loop separately, for open strings stretched between specific representatives of the orbifold orbit,
 and then the sum is performed at the end. 
 The localized states ($\N=1$ sectors)
 that we discuss below arise at the intersections of these orbifold (and orientifold) images of D-branes. 
 
 Once we have identified under the orbifold action, a brane may still self-intersect in the actual orbifold space at nonzero intersection angle. However as we have seen in this paragraph,  what is generated 
 as factors in the annulus amplitudes are the intersection numbers in the covering torus, as opposed to intersection numbers in the actual orbifold space. This will  hopefully be clear in the calculation below,
 and in figure \ref{braneimage}. 
 
There is another potentially slightly confusing point here: as argued earlier, our vertex operators  are adapted to a brane at a specific angle. But
  the invariant (orbifold-neutral) open-string states consist of superpositions of open strings stretched between representatives of the orbifold orbit, i.e.\ we need to use many different angles. This is consistent because in the covering space, the D-brane moduli of each image brane are independent, and only when the superposition is formed to make an orbifold-neutral state do we get a single set of D-brane moduli. This is in fact the same logic given above for states running in the loop: if the external-state  D-brane moduli are viewed as independent charged states, the invariant states can be formed  at the very end of the calculation by summing over charged states, i.e.\ effectively applying a projector. 
    
For the M\"obius strip the situation is more complicated, as also $k\neq 0$ sectors contribute,
i.e. now the open string states in the loop can be rotated while traversing the loop. Similar 
to the annulus, we can write the M\"obius amplitude in the open string channel as a trace
\be
\sum_k \sum_{a,a'} \langle a,a' | q^{H} \Omega \cR \twist^k | a,a' \rangle\ ,
\ee
where $a' \in [a]$. In order to see which amplitudes actually contribute, let us consider the two 
cases $a' = \cR \twist^m a$ and $a' = \twist^m a$ separately. In the first case we have 
\be
\langle a,  \cR \twist^m a| q^{H} \Omega \cR \twist^k | a, \cR \twist^m a \rangle &=& 
\langle a,  \cR \twist^m a| q^{H} \Omega | \cR \twist^k a, \cR \twist^k \cR \twist^m a \rangle \\
&=& \langle a,  \cR \twist^m a| q^{H} \Omega | \cR \twist^k a, \twist^{m-k} a \rangle \\
& = & \langle a,  \cR \twist^m a| q^{H} | \twist^{m-k} a , \cR \twist^k a  \rangle \ , \label{moebius1}
\ee
where we used 
\be \label{relationrtwist}
\twist^k \cR = \cR \twist^{N-k}
\ee
when going from the first to the second line. It is obvious from \eqref{moebius1} that there will only be a non-vanishing 
contribution to the trace if $m=k$ or if $m=k+N/2$. The latter does not contribute, but it is not obvious from this argument  --- see \cite{Lust:2003ky}.

In the second case we have instead
\be
\langle a, \twist^m a| q^{H} \Omega \cR \twist^k | a, \twist^m a \rangle &=& 
\langle a,  \twist^m a| q^{H} \Omega |  \cR \twist^{k} a, \cR \twist^{k+m} a \rangle \\
& = & 
\langle a, \twist^m a| q^{H} | \cR \twist^{k+m}  a, \cR \twist^{k} a \rangle \ . \label{moebius2}
\ee
We now see (again using \eqref{relationrtwist}) that the necessary condition for a non-vanishing contribution is 
\be
a =  \cR \twist^{k+m}  a\ .
\ee 
In other words, we need $a$ to lie on top of the $O_{k+m}$ orientifold plane. If we assume that none of the 
branes lies on top of the orientifold planes (along all three tori), 
we obtain the result that the non-vanishing contributions from the M\"obius amplitudes are
\be  \label{moebiusampl}
\sum_k \sum_{a} \langle a, \cR \twist^k a | q^{H} \Omega \cR \twist^k | a, \cR \twist^k a \rangle\ .
\ee
It was shown in \cite{Lust:2003ky} that these amplitudes have the feature that 
in the closed string channel only untwisted closed strings are exchanged. 
It is still possible that $a$ and $\cR \twist^k a$ are parallel along a single torus 
(i.e.\ $a$ lies on top of $O_k$ along this particular torus), in which case the contribution would preserve 
$\N=2$ supersymmetry. 

In the following section, we will calculate the annulus and M\"obius amplitudes in turn. 




\section{$\N=1$ supersymmetric sector}
\label{n1}

In this section we consider the open strings stretched between two stacks of D6-branes intersecting at non-vanishing angles 
along every internal torus,
where the sum of the three angles is zero (modulo $\pi$):
\be
\varphi_1+\varphi_2 + \varphi_3 = 0 \; , \quad \varphi_1, \varphi_2, \varphi_3 \neq 0 \; .
\ee
This configuration preserves $\N=1$
supersymmetry in four dimensions and contributions
due to these strings are sometimes called $\N=1$ sector contributions (cf. for example \cite{Lust:2003ky}).
It is important not to confuse this $\N=1$ untwisted sector with $\twist^k$-twisted sectors,
which, for $k\neq N/2$ and in type IIB, are also called $\N=1$ sectors.

Let us mention that the angles $\varphi$ are related to the $\theta$ used until now by the simple relation
\be
\varphi = \frac{\pi}{2} - \theta\ ,
\ee
cf.\ figs.\ \ref{Xcoords} and \ref{tiltedtorus}.


\subsection{Annulus amplitude}
We proceed to 
calculate the annulus amplitude in the $k=0$ sector (the untwisted sector), 
since as argued above this is the only sector that can contribute.   
In particular, for concreteness, we 
compute the 2-point function of open string scalars 
$\Phi_3$ and ${\bar\Phi}_{\bar{3}}$ polarized along the third two-torus and belonging to the brane (stack) $a$. 
For the complex annulus coordinate $\nu_\A$ (cf.\ \eqref{nuw}),
the vertex operators are integrated along the positive imaginary axis
from the origin to $\tau_\A=it/2$. 
Using the vertex operator \Ref{Vtotal}, 
the expression is 
(see e.g.\ \cite{Berg:2004sj}) 
\be \label{A_66_general}
\langle \Phi_3 \bar \Phi_{\bar{3}} \rangle_{\A} &=& 
\frac{1}{4N}  \int_0^{i \infty} d\tau_{\A} \int_0^{\tau_{\A}} \! d\nu_\A
 \sum_{{\rm images}}  \sum_{{\alpha\beta}\atop{\rm even}}  \eta_{\alpha,\beta} 
 \cz_{\A}^{\rm tot} \zba{\alpha}{\beta} \; 
\langle V_{\Phi}(\nu_\A)V_{\bar{\Phi}}(0) \rangle_{\A}^{\alpha, \beta}   \\
&=& 
\delta \xi e_3 \bar{e}_{\bar{3}} \int_0^\infty dt \int_0^{t/2} d\nu  \sum_{{\rm images}} 
 \sum_{{\alpha\beta}\atop{\rm even}} \tr ( \lambda_1 \lambda^\dag_2) \tr(\gamma_6^0) \\
&& \hspace{0cm}
 \times  
  \eta_{\alpha,\beta} 
 \cz^{\rm ext}_\A\zba{\alpha}{\beta} \cz_{\A}^{\rm int} \zba{\alpha}{\beta} 
e^{-\delta\langle X(i \nu) X(0)\rangle}
\langle \Psi(i \nu) \bar \Psi(0) \rangle_{\A}^{\alpha, \beta}
\langle \psi(i \nu) \psi(0) \rangle_{\A}^{\alpha, \beta} \ ,
\nonumber
\ee
where we wrote $\cz_{\A}^{\rm tot} = \cz_{\A}^{\rm ext}\cz_{\A}^{\rm int}$, 
$\cz_{\A}^{{\rm ext}}$ is the spacetime annulus partition function \Ref{Z4},
$\cz_{\A}^{{\rm int}}$ the internal annulus partition function from \Ref{Aab2} 
and $\nu_\A = i \nu$ for $\nu$ real. Also, 
the normalization is
\be \label{xi}
\xi = -\frac{g_o^2 \alpha'}{8 N L_3^2}\ ,
\ee
where $N$ is the order of the orientifold group, and $\delta = p_1 \cdot p_2$. On-shell, this would vanish. However, in order to calculate the correction to the K\"ahler metric, one can relax momentum conservation artificially and read off the metric as the coefficient of $\delta$. A similar procedure was used often before, see for instance \cite{Atick:1987gy,Minahan:1987ha,Poppitz:1998dj,Bain:2000fb,Antoniadis:2002cs,Berg:2005ja,Benakli:2008ub,Anastasopoulos:2011kr}.
\begin{figure}
\begin{center}
\includegraphics[width=0.3\textwidth]{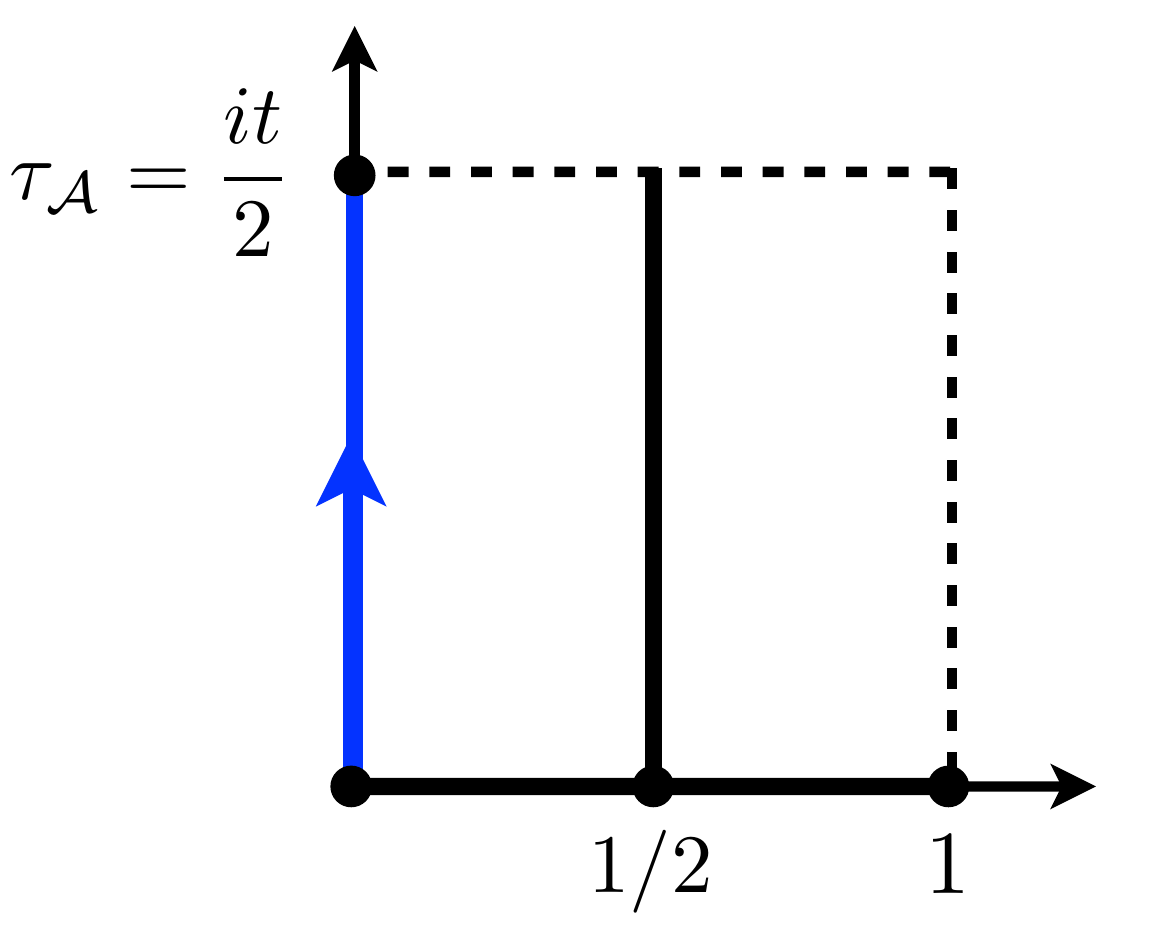}
\caption{
Integration region for $\nu_\A$.}
\end{center}
\end{figure}
We can now insert the expressions
for the worldsheet correlators from appendix \ref{wscorr} and perform the traces
over the $U(1)$ subgroups in which our Chan-Paton factors sit (the matrices $\lambda$ are diagonal and have $N_a$ entries of $1$, at positions which are appropriate for the member of the orientifold orbit). 
\be \label{A_66}
\langle \Phi_3 \bar{\Phi}_{\bar{3}} \rangle_{\A}
&=& \delta \xi e_3 \bar{e}_{\bar{3}}
 \int_0^{\infty} \frac{dt}{(4\pi^2 \alpha' t)^2} \int_0^{t/2} d\nu \, 
R_{\delta} \\
&& \times  \sum_{{\rm images}} N_a N_b\sum_{\alpha,\beta= {\rm even}} \eta_{\alpha,\beta}   \cz^{\rm ext}_\A\zba{\alpha}{\beta}(\tau_{\A}) 
\cz_{\A}^{{\rm int}}\zba{\alpha}{\beta}(\tau_{\A}) 
G_F\zba{\alpha}{\beta}(i \nu, \tau_{\A}) G_F\zba{\alpha+v}{\beta}(i \nu, \tau_{\A})\ , \nonumber 
\ee
where $G_F\zba{\alpha}{\beta}(i \nu,\tau_{\A})$ is
the fermionic correlator \Ref{GFtheta},
 $v$ is
\be
 v\equiv  v^{3}_{ab} = {1 \over \pi} (\varphi^3_a-\varphi^3_b)
\ee
as defined in the appendix in equation \Ref{vab},
where we note that this depends on the representatives $a$ and $b$, which
are not indicated explicitly. 
Finally the  function $R_{\delta}$ is, from \Ref{boscorr},
\be \label{R_exp}
R_{\delta}(\nu,t)= e^{-\delta \langle X(i\nu)X(0) \rangle_{\A}} = \left|{ \tht_1(i \nu,\tau_{\A}) \over
\tht_1'(0,\tau_{\A}) } \right|^{2 \alpha' \delta} \! e^{-{4 \pi \alpha' \delta  \nu^2  \over t }} \ .
\ee
We have left the sum over brane images implicit in \Ref{A_66}, as it 
requires some more notation that we will not need in this section, 
and we relegate the details to the appendix. 

We first perform the spin structure sum over $\alpha,\beta$, using the quartic Riemann identity
\be
\sum_{\alpha,\beta=0,1/2 \atop {\rm even}} \eta_{\alpha, \beta} 
\thba{\alpha}{\beta}(i \nu, \tau)\thba{\alpha+v^{3}}{\beta}(i \nu, \tau) 
\prod_{i=1,2} \thba{\alpha+v^i}{\beta}(0, \tau) \nonumber \\
=\thba{1/2}{1/2}(i \nu, \tau)\thba{1/2+v^{3}}{1/2}(i \nu, \tau) 
\prod_{i=1,2} \thba{1/2+v^i}{1/2}(0, \tau)\ .
\ee
Doing so, the amplitude reduces to
\be \label{A_66_b}
\langle \Phi_3 \bar \Phi_{\bar{3}} \rangle_{\A}
&=&\delta  \xi e_3 \bar{e}_{\bar{3}} 
 \sum_{{\rm images}} N_a N_b \prod_{i=1}^3 I^i_{ab} \int_0^{\infty} \frac{dt}{(4\pi^2 \alpha' t)^2} \int_0^{t/2} d\nu 
 R_\delta(\nu,t) G_F\zba{1/2+v^{3}}{1/2}(i \nu, \tau_\A)\ . \nonumber 
\ee
The function $R_{\delta}$ acts as an infrared regulator for $\delta=p_1 \cdot p_2 \rightarrow 0$,
but we will argue that it does not contribute and in fact we can set $R_{\delta}\equiv 1$
in the $\delta \rightarrow 0$ limit.

We have now reduced the calculation to computing the integral
\be \label{Integral} 
\ci = \int_0^{\infty} \frac{dt}{t^2} \int_0^{t/2} d\nu \, 
R_{\delta}(\nu,t) G_F\zba{1/2+v^{3}}{1/2}(i \nu, \tau_\A)\ .
\ee
It will be convenient to immediately transform to the closed string channel, with
$\ell=1/t, \tilde{\nu}\equiv 2 \nu \ell$
\be \label{GFtree0} 
\ci =  - i \int_0^{\infty} d\ell \int_0^1 d \tilde{\nu} \,  
\tilde{R}_{\delta}(\tilde{\nu},\ell) G_F\zba{1/2}{1/2 + v^{3}} (\tilde{\nu}, 2 i \ell) \ ,
\ee
where
\be \label{R_exp2}
\tilde{R}_{\delta}(\tilde{\nu}, \ell)= \left|{\tht_1(\tilde{\nu},2i\ell) \over
 2\ell\, \tht_1'(0,2i\ell) } \right|^{\alpha' \delta}  \ .
\ee
Here we performed the modular S transformations \Ref{XXS} and \Ref{GFS}.

This integral is divergent in several regions of the space of worldsheet moduli $\tilde{\nu}$ and $\ell$. 
There is the usual tadpole divergence for $\ell \rightarrow \infty$ and a possible
divergence at $\ell \rightarrow 0$, that we regulate by cutoffs:
\be \label{GFtree} 
\ci =  - i \int_{\mu}^{\Lambda} d \ell \int_0^1 d \tilde{\nu} \,  
\tilde{R}_{\delta}(\tilde{\nu},\ell) G_F\zba{1/2}{1/2 - v^{3}} (\tilde{\nu}, 2 i \ell) \ .
\ee
The $\ell \rightarrow \infty$ divergences cancel between diagrams using tadpole cancellation conditions from the vacuum amplitude, as we go through in detail below. 

There is also a potential divergence from vertex operator collisions $\tilde{\nu}\rightarrow 0$,
which is regulated by keeping a nonzero (but infinitesimal) $\delta$.  We now proceed to show that this is cancelled for each diagram separately. 


\subsection{Vertex collision divergence}   \label{infrared_annulus}

We are interested in the $\delta \rightarrow 0$ limit of 
\be \label{openint}
&& \ci_{\nu} =\int_0^1 d \tilde{\nu} \,  
\tilde{R}_{\delta}(\tilde{\nu},\ell) G_F\zba{1/2}{1/2 + v^{3}} (\tilde{\nu}, 2 i \ell)  \; . 
\ee
The reason that we cannot simply immediately set $\delta = 0$,
thereby removing $\tilde{R}_\delta$ altogether, is that there 
is a vertex collision pole at $\tilde{\nu}=0$ and at $\tilde{\nu}=1$. For instance at $\nu=0$ we have:
\be  \label{limit_G_R_1}
G_F(\tilde{\nu})  \rightarrow {1 \over \tilde{\nu}} \; , \quad
\tilde{R}_{\delta}(\tilde{\nu}) \rightarrow \tilde{\nu}^{\delta} \qquad
\mbox{as } \tilde{\nu}\rightarrow 0 \; ,  
\ee
where we absorbed $\alpha'$ into $\delta$ in \Ref{limit_G_R_1} i.e. $\alpha' \delta \to \delta$.
Because $\tilde{\nu}^{-1+\delta}$ is integrable
for nonzero positive $\delta$, we see that in fact $\tilde{R}_{\delta}$ regulates the
integral in the $\tilde{\nu}\rightarrow 0$ limit. 
(It may be useful to recall that the limit
$\delta\rightarrow 0$ is the {\it long-distance} limit in spacetime, but there is some interplay with a short-distance singularity on the worldsheet.)
But because of
\be
\lim_{\tilde{\nu} \to 1} G_F\zba{\alpha}{\beta}(\tilde{\nu}, 2 i \ell) 
&=& \lim_{\tilde{\nu} \to 1} -e^{2 \pi i \alpha} G_F\zba{\alpha}{\beta}\left(\tilde{\nu}-1, 2 i \ell \right) =\frac{-e^{2 \pi i \alpha}}{\tilde{\nu}-1} 
\stackrel{\alpha=1/2}{\longrightarrow}{1 \over \tilde{\nu}-1}
\ee
there is also a corresponding pole at $\tilde{\nu}=1$:
\be
 \label{limit_G_R_2}
G_F(\tilde{\nu}) \rightarrow {1 \over\tilde{\nu}-1} \; , \quad
\tilde{R}_{\delta}(\tilde{\nu}) \rightarrow (\tilde{\nu} - 1)^{\delta} \qquad
\mbox{as } \tilde{\nu}\rightarrow 1\; .
\ee
We see that $\tilde R_{\delta}$ regulates the divergences at both poles of $G_F$. 
We would now like to show that the divergences actually cancel each other,
and we do so by subtracting a function of suitable periodicity and singularity,
which turns out to be the cotangent function (as one can easily
see also from equation \Ref{GFWW} below).
We split the integrand as follows:
\be \label{splitI}
\ci_{\nu} = \ci_{\nu,1} + \ci_{\nu,2} = 
 \int_0^1 d \tilde{\nu}\, \tilde{R}_{\delta}(\tilde{\nu}) \left( G_F(\tilde{\nu}) - \pi \cot \pi \tilde{\nu} \right) 
+ \int_0^1 d \tilde{\nu}\, \tilde{R}_{\delta}(\tilde{\nu}) \pi \cot \pi \tilde{\nu}\ .
\ee 
Having subtracted the poles, we will be able to
take the $\delta\rightarrow 0$ limit of the first integral
which will be our potentially finite contribution.
In the second integral we have a potential
 divergence from each pole as we let $\delta\rightarrow 0$. However, we need only observe that
 for any nonzero $\delta$, no matter how small,
 \be
 \ci_{\nu,2} = 
 \int_0^1 d \tilde{\nu}\, \tilde{R}_{\delta}(\tilde{\nu}) \pi \cot \pi \tilde{\nu} =0\ ,
 \ee
 because the regulating function is even under reflection at $\tilde{\nu}=1/2$: $\tilde{R}_\delta(1-\tilde{\nu})=\tilde{R}_\delta(\tilde{\nu})$, and
 the cotangent is odd: $\cot(\pi(1-\tilde{\nu}))=-\cot(\pi\tilde{\nu})$. 
 What happens is simply that the two poles,
 the one at $\tilde{\nu}=0$ and the one at $\tilde{\nu}=1$, cancel each other. 
 We let this be our regularization prescription, i.e.\ in principle
 we keep a nonzero $\delta$, but we may make it arbitrarily small
 such that it will not affect our results.  
 \begin{figure}
 \begin{center}
 \includegraphics[width=0.4\textwidth]{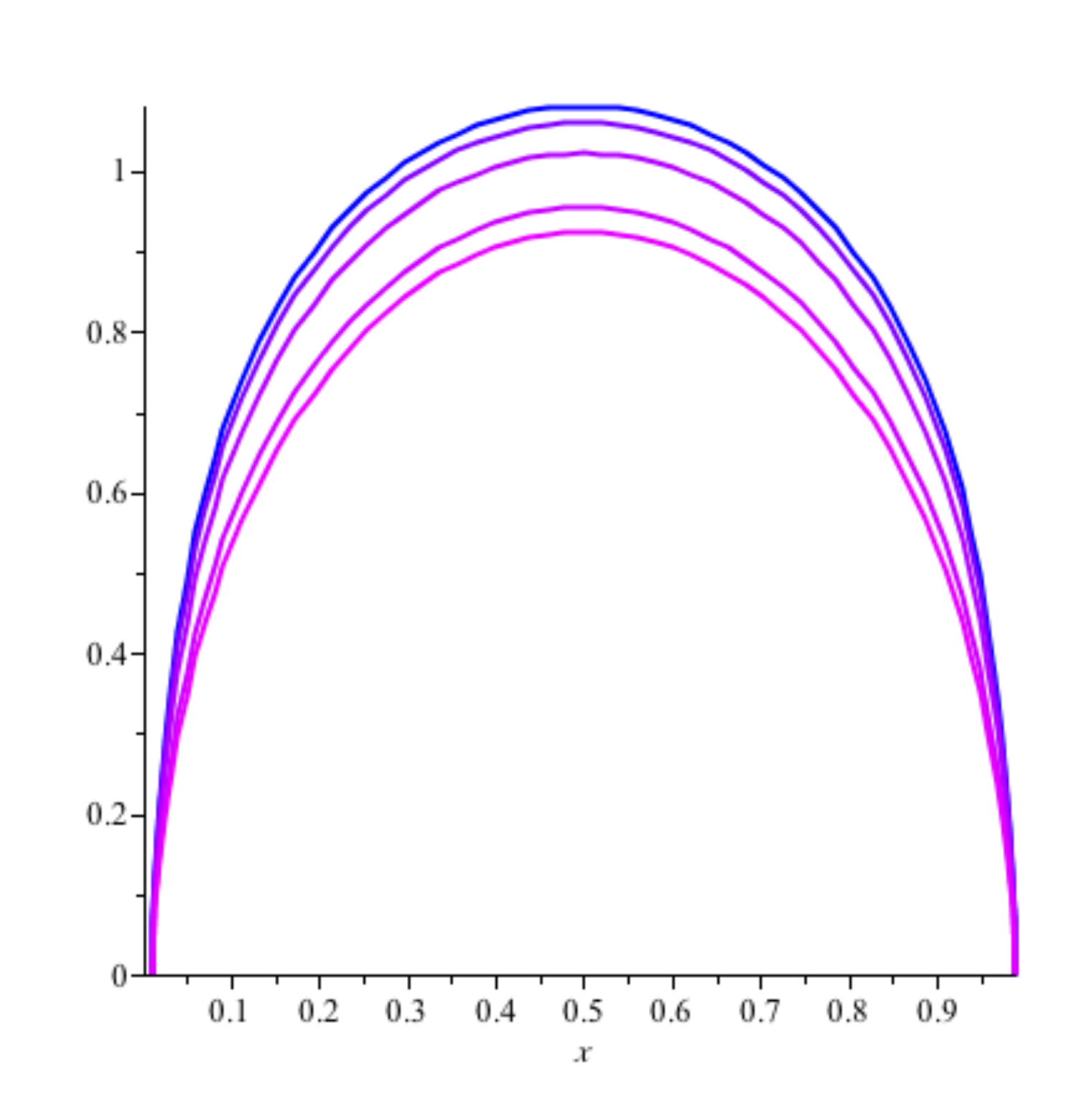}
\hspace{-5mm}
\caption{A plot of $\int_0^x \tilde R_{\delta}(\tilde \nu) \cot (\pi \tilde \nu) d \tilde \nu$ for $x=0..1$, $\ell=0.4$, $v=1/3$, 
$\delta=[0.01,0.02,0.04,0.08, 0.1]$.
For $x=1$, the integral yields zero for all nonzero values of $\delta$.}
\end{center}
\hspace{-5mm}
\end{figure}
We note that a similar argument was put forward in
\cite{Anastasopoulos:2011kr}.
 
The conclusion is that we may safely set $\tilde{R}_{\delta}\equiv 1$ 
in $\ci_{\nu,1}$ for the remainder of this discussion.

Although it is not quite obvious at this point, we will be able to make a very similar argument for
the M\"obius strip amplitude.


\subsection{Tadpole cancellation}

As part of our 
quest to compute (\ref{GFtree}), we will
now analyze the  closed-string IR behavior ($l \to \infty$) of the integrand.  
This is the region that would exhibit divergences in the vacuum amplitude
if tadpoles were not cancelled, so we expect that it will be cancelled between diagrams
if we assume that the brane configuration cancels tadpoles. We now outline this calculation for the two-point function, with most of the detail given in the appendix. 

Using the product representation of the theta functions in (\ref{GFtheta}) it is easy to see
that 
\be \label{annulus_UV}
G_F\zba{1/2}{1/2+v^{3}}(\tilde{\nu}, 2 i \ell) \; \stackrel{\ell \to \infty}{\rightarrow} \; 
\frac{\pi \sin\pi(\tilde{\nu}+v^{3})}{\sin\pi v^{3} \sin\pi \tilde{\nu}}
=\pi \cot\pi \tilde{\nu}+\pi \cot\pi v^{3}.
\ee
The first term is familiar from our discussion of the vertex operator collision divergence, the
second term depends on the angle, which in turn
depends on the intersection numbers. 

First it is useful to recall
 the  $\ell \rightarrow \infty$ divergences of the 2-point function of {\it vectors}, 
 as opposed to D-brane scalars.  As is well known, an efficient 
 way to compute threshold corrections to gauge couplings is to consider the 
vacuum amplitude deformed by an external background
field $B$ and expanded to order $B^2$. In \cite{Lust:2003ky}
it was shown that 
\be  \label{I3}
\mbox{NSNS tadpole for {\it vectors}  }\quad \propto \quad I^3 \sum_{i=1}^3 \cot \pi v^i \times \mbox{(regulated divergence)}\ .
\ee
Because $\cot \pi v^i = V^i/I^i$ (see \eqref{cotv}),
this becomes of the schematic form ``$I^2 V$''. This is denoted $\kappa$ in \cite{Lust:2003ky}
and it is shown that the 
detailed expressions for $\kappa$ cancel for the explicit example of $T^6/(\mathbb{Z}_2 \times \mathbb{Z}_2)$.

In our calculation of the D-brane scalar 2-point function,
we obtain a similar coefficient but without the sum over the three 2-tori:
\be
\langle \Phi_3 \bar \Phi_{\bar{3}} \rangle^{\rm UV}= 
I^3 \cot \pi v^3 \times \mbox{(regulated divergence)} \ .
\ee
This is also of the schematic form ``$I^2 V$'',
and it vanishes by vacuum tadpole cancellation,
as we show explicitly for $T^6/\mathbb{Z}_6'$ in appendix \ref{app:tadpole}. 
Note, however, that this is a somewhat stronger
result than the result that \Ref{I3} vanishes {\it after} summing over the three two-tori. 

In addition, we have the first term in \eqref{annulus_UV}, which is a possible divergence that has no analog in the 
background field calculation, the divergence from vertex operator collisions: 
\be
I^3 \cot \pi \tilde{\nu} \qquad \mbox{for $\tilde{\nu} \rightarrow 0$ or $1$}\ .
\ee
We have already shown in the previous section that this kind of divergence cancels
{\it before} taking the $\ell\rightarrow \infty$ limit, and for each diagram separately,
when we keep a nonzero $\delta$. (We showed this for the annulus diagrams and will show it 
for the M\"obius diagrams in sec.\ \ref{Mstrip}.) However, it is somewhat useful to also 
exhibit that in fact this divergence would also cancel between diagrams,
without using details of the integrand. This does require some actual 
model-dependent calculation, which the argument of cancellation in the integrand
does not, so there is a certain complementarity of these two discussions. 

Indeed, as indicated above, since the integrand is  independent of the angles $\varphi_{ab}$, the sum will be of the schematic form $I^3$.
The contribution to the coefficient of the $\tilde \nu$-integral from some worldsheets in some sectors are nonzero.
But it is easy to check that the total contributions to the coefficient vanish for any brane configurations,
so this is automatically zero. More details are given at the end of appendix \ref{divcancel}. 

It now remains to calculate the finite contribution from the first term in \eqref{splitI}.


\subsection{Vanishing of UV-finite contribution}

We want to compute the finite integral
\be \label{UV-finite} 
\ci_{\rm finite} = - i \int_{\mu}^{\Lambda} d \ell \int_0^1 d \tilde{\nu} \,  
\left( G_F\zba{1/2}{1/2 + v^{3}} (\tilde{\nu}, 2 i \ell) - \pi \cot\pi \tilde{\nu} - \pi \cot\pi v^{3} \right). 
\ee
Since we have argued that closed-string infrared $(\ell \rightarrow \infty)$ divergences
cancel between diagrams due t tadpole cancellation, we could in principle remove the cutoff $\Lambda$ from \eqref{GFtree},
but we will keep it as the integrals are still divergent for each diagram separately. 
We also have the explicit $\ell\rightarrow 0$ cutoff 
$\mu$. It will be easy to see that none of our results for the finite parts depend on these regulators.
For the integrand of \Ref{UV-finite}, it is particularly convenient
to use the representation   
\be \label{GFWW}
G_F\zba{1/2}{1/2 + v^{3}} (\tilde{\nu}, 2 i \ell) = \pi \cot\pi \tilde{\nu} + \pi \cot\pi v^3 
+ 4 \pi \sum_{m,n=1}^{\infty}e^{-4 \pi \ell m n} \sin(2\pi n \tilde{\nu}+2\pi m v^3),
\ee
of the fermionic Green's function, cf.\ \eqref{target}.
Since this representation is perhaps not familiar to all readers, in appendix \ref{WW}
we provide an elementary proof that it is equivalent
to the representation \Ref{GFtheta} in terms of Jacobi theta functions.
We then see that
(\ref{UV-finite}) is nothing but
\be  \label{vanishing}
\ci_{\rm finite} 
&=& - 4 \pi i  \int_{\mu}^{\Lambda}d\ell \int_0^1 d\tilde{\nu} 
\sum_{m,n=1}^{\infty}e^{-4 \pi \ell m n} \sin(2\pi n \tilde{\nu}+2\pi m v^3) \nonumber \\
&=& - 4 \pi i \int_{\mu}^{\Lambda}  d\ell
\sum_{m,n=1}^{\infty}e^{-4 \pi \ell m n} \int_0^1 d\tilde{\nu} \sin(2\pi n \tilde{\nu}+2\pi m v^3) \nonumber \\
&=&  4 \pi i\int_{\mu}^{\Lambda}  d\ell
\sum_{m,n=1}^{\infty}e^{-4 \pi \ell m n} \times \left[ \frac{\cos(2\pi n \tilde{\nu}+2\pi m v^3)}{2\pi n} 
\right]_{\tilde{\nu}=0}^{\tilde{\nu}=1} \nonumber \\
&=&  4 \pi i \int_{\mu}^{\Lambda}  d\ell
\sum_{m,n=1}^{\infty}e^{-4 \pi \ell m n} \times 0 \nonumber \\
&=& 0. 
\ee
The integration over vertex position $\tilde{\nu}$ gives zero, 
by periodicity. Note
that if we had not put the UV cutoff $\mu$, 
the contribution would naively have diverged at the $\ell=0$ end (where
the exponential in the sum over $m$ and $n$ becomes 1). We see that the result 
vanishes for any finite value of $\mu$ and $\Lambda$ and, thus, also in the limit $\mu, 1/\Lambda \rightarrow 0$. 

In appendix \ref{app:vanishing}, we prove this result in a quicker and less rigorous way
by contour integration. 


\subsection{M\"obius strip amplitude}
\label{Mstrip}

Now we consider the M\"obius strip amplitude, describing an open string starting on a brane $a$ and ending on one of its orientifold images, cf.\ \eqref{moebiusampl}.
The orientifold planes of $\mathbb{Z}_6'$ are given explicitly in the appendix;
 there are six distinct orientifold planes $O_k$ for $k=1,\ldots , 6$. We assume that the brane $a$ along every torus does not sit on any orientifold plane, 
so that brane $a$ and its orientifold images have non-vanishing intersection angles.

\begin{figure}
\begin{center}
\includegraphics[width=0.4\textwidth]{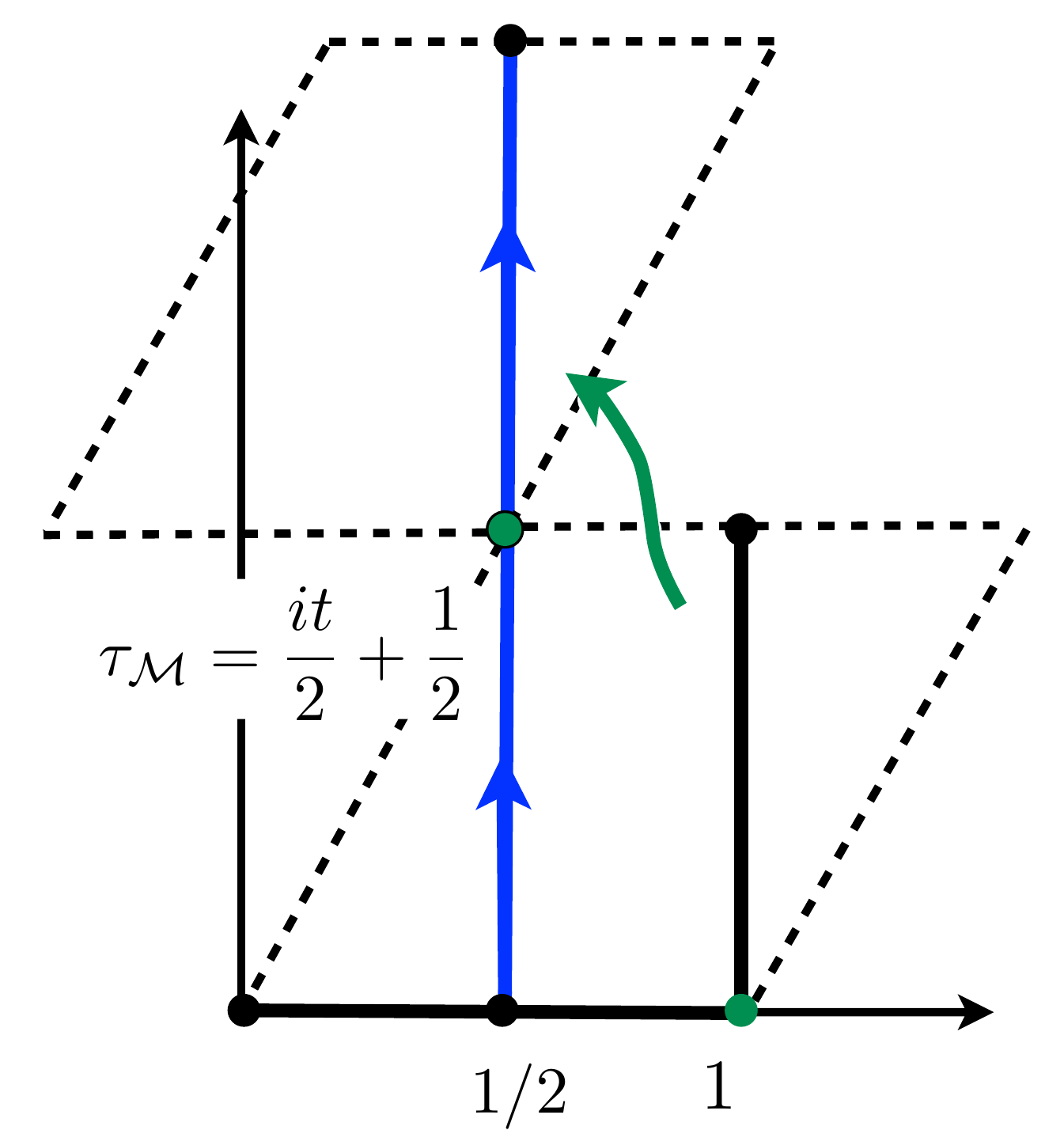}
\caption{
Integration region for $\nu_\M$.}
\end{center}
\end{figure}
Similarly to above (cf.\ \eqref{A_66_general} and \eqref{A_66}),
\be 
\langle \Phi_3 \bar \Phi_{\bar{3}} \rangle_{\M} &=& 
- \frac{\delta}{4N}  \int_{1/2}^{i \infty + 1/2} d\tau_\M \int_{1/2}^{it+1/2} \! d\nu_\M 
\sum_{{\rm images}} \sum_{k=0}^{N-1}  \sum_{{\alpha\beta}\atop{\rm even}} 
\eta_{\alpha,\beta} \cz_{\M,k}^{\rm tot} \zba{\alpha}{\beta} \; 
\langle V_{\Phi}(\nu_\M)V_{\bar{\Phi}}(1/2) \rangle_{\M}^{\alpha, \beta}    \\
&=& -\delta e_3 \bar{e}_{\bar{3}} \xi \, 
\int_0^{\infty}  \frac{dt}{(4\pi^2\alpha' t)^2} \int_0^{t} d\nu \, 
R_{\delta}(\nu,t) \sum_{{\rm images}} N_a  \sum_{k=0}^{N-1} \rho_k
\label{M_66} \\
&&\times \sum_{\alpha,\beta= {\rm even}} \eta_{\alpha,\beta} 
 \cz^{{\rm ext}}\zba{\alpha}{\beta}(\tau_\M) 
 \cz^{{\rm int, k}}\zba{\alpha}{\beta}(\tau_\M) 
G_F\zba{\alpha}{\beta}(i \nu, \tau_\M) G_F\zba{\alpha+ 2 v_{\rm O}}{\beta-v_{\rm O}}(i \nu, \tau_\M)\ , \nonumber 
\ee
where $\tau_\M={i t \over 2} + {1 \over 2}$, $\nu_\M = i \nu + 1/2$, $v_{\rm O}$ is
\be
 v_{O} \equiv v^{3}_{a,O_k}= -{1 \over \pi} (\varphi^3_a-\varphi^3_{O_k})
\ee
as defined in the appendix in equation \Ref{vOab},
and $\xi$ is the same as for the annulus, cf.\ \eqref{xi}. 
Note that $v_{\rm O}$ depends on the sector $k$. 
The external spacetime partition function $\cz^{\rm ext}$ is the same as (\ref{Z4}),
the internal partition function $\cz^{{\rm int,k}}$ 
is given in \Ref{Z_int_M}.
The phase $\rho_k$ arises from the Chan-Paton matrices representing the twist action 
$\Omega \cR \twist^k$ on the branes (see the remarks below (2.11) in \cite{Lust:2003ky}).  
Note that the angle $v_O$ 
is that between the brane and the orientifold plane, which is half
the angle between the brane and its orientifold image. 

After summation over even spin structures using the quartic Riemann identity 
\be  \label{quarticR}
\sum_{\alpha,\beta= {\rm even}} \eta_{\alpha,\beta} \thba{\alpha}{\beta}{(\nu,\tau)} 
\thba{\alpha+h_3}{\beta+g_3}{(\nu,\tau)} \prod_{i=1}^2 \thba{\alpha+h_i}{\beta+g_i}{(0,\tau)}
= \thba{1/2}{1/2}{(\nu,\tau)} \thba{1/2+h_3}{1/2+g_3}{(\nu,\tau)} 
\prod_{i=1}^2 \thba{1/2+h_i}{1/2+g_i}{(0,\tau)}\nonumber \\
\ee
with $\sum_{i=1}^3 h_i=0=\sum_{i=1}^3 g_i$, the integral of (\ref{M_66}) 
reduces to
\be \label{Integral_M_R} 
\ci^\M = \int_0^{\infty} \frac{dt}{t^2} \int_0^{t} d\nu \, R_{\delta}(\nu,t) G_F\zba{1/2+2 v_{\rm O}}{1/2 - v_{\rm O}}(i \nu, \tau_\M)\ .
\ee
Note that $R_{\delta}(\nu,t)$ here is not the same as $R_{\delta}(\nu,t)$ for the annulus, but it is defined analogously, cf.\ eq.\ \eqref{R_exp}. The explicit form does not play any role. The correlator can be rewritten in terms of the tree channel:
\be 
G_F\zba{1/2 + 2 v_{\rm O}}{1/2- v_{\rm O}}\left(i \nu, \tau_{\M} \right) &\stackrel{\tilde{\nu} \equiv 4 \nu \ell}{=} & - 4 i \ell  G_F\zba{1/2}{1/2 + v_{\rm O}}\left( \tilde{\nu}, 
\ell_{\M} \right)\ , 
\ee
where we performed the sequence $ST^2S$ of modular transformations: 
\be
\tau_{\M}= {i t  \over 2} + \frac{1}{2} \;\to\; -\frac{1}{\tau_{\M}} \;\to\; 
- \frac{1}{\tau_{\M}} +2  \;\to\; \left(\frac{1}{\tau_{\M}} - 2  \right)^{-1} =  2 i \ell- \frac{1}{2}
=: \ell_{\M} \ .
\ee 
In the last step, we used the relation (see for instance \cite{Antoniadis:1999ge})
\be
t = \frac{1}{4 \ell}\ .
\ee 
Thus, in the tree channel the amplitude is
\be \label{M_tree}
- 4 i \int_0^{\infty} d\ell \int_0^{1} d\tilde{\nu} \,   \tilde{R}_\delta(\tilde{\nu},\ell)
G_F\zba{1/2}{1/2 + v_{\rm O}}\left(\tilde{\nu}, \ell_{\M}\right)\ .
\ee 
We see that since this is very similar to the annulus closed channel amplitude,
in particular the only non-half-integer characteristic of $G_F$ is the lower one,
the same argument for  cancellation of infrared divergences goes through and we will set 
$\tilde{R}_\delta\equiv 1$. 
The IR behavior of the integrand can again be isolated,
\be
G_F\zba{1/2}{1/2+v_{\rm O}}\left(\tilde{\nu}, \ell_{\M}\right) \stackrel{\ell \to \infty}{\rightarrow} 
\frac{\pi \sin(\pi\tilde{\nu}+\pi v_{\rm O})}{\sin(\pi v_{\rm O}) \sin(\pi \tilde{\nu})}
=\pi \cot(\pi \tilde{\nu})+\pi \cot(\pi v_{\rm O})\ ,
\ee
and the remaining finite part is
\be \label{M_UV_finite}
\ci_{\rm finite}^\M = 
- 4i \int_0^{\infty} d\ell \int_0^{1} d\tilde{\nu} \, 
\left[G_F\zba{1/2}{1/2 +v_{\rm O}}\left(\tilde{\nu}, \ell_{\M}\right)
-\pi \cot(\pi \tilde{\nu})-\pi \cot(\pi v_{\rm O})\right].
\ee
Using a similar representation to the one that we used for the annulus amplitude
\be
G_F\zba{1/2}{1/2 + v} \left( \tilde{\nu}, \ell_{\M}\right)&=&
 \pi \cot (\pi v) + \pi \cot (\pi \tilde{\nu}) \nonumber \\
&& + 4 \pi \sum_{m,n=1}^{\infty} \left(-e^{- 4 \pi \ell}\right)^{m n}
\sin\left(2 \pi n  \tilde{\nu}+ 2 \pi  m  v \right)
\ee
and following similar steps as in the case of the annulus, we find
\be
\ci_{\rm finite}^\M = 0 \; . 
\ee  
Neither annulus nor M\"obius $\N=1$ amplitudes contribute finite parts
to the two-point function,
and we have shown that the divergent parts cancel,
so there are no contributions at all from
these sectors. 


\section{$\N=2$ supersymmetric sector}
\label{n2}

In this section we investigate the cases where two branes are parallel along internal tori. 
The supersymmetry condition $\sum_{j=1}^3 v^j=0$ for annulus ($\sum_{j=1}^3 v^{k;j}=0$ for M\"obius) 
requires that two branes have vanishing angle along at most one torus. 
This sector is a so called $\N=2$ sector (cf. \cite{Lust:2003ky}).  
The partition functions can be obtained using \Ref{Aab2}, \Ref{Z_int_M}, \Ref{Amod} and \Ref{Mmod} in the appendix. 
The correlators remain the same as for the $\N=1$ sectors.
Thus the spin structure dependence of the amplitudes is the same as in the case of $\N=1$ sectors, 
and the spin structure summation proceeds in the same way
using \eqref{quarticR}. 
If the $i$th torus has vanishing intersecting angle, that is, if $h_i=0=g_i$ for $i=1$ or 2, then the RHS is zero.  
Thus it follows that if two branes are parallel along either the first torus or the second torus 
then the spin-structure sum gives zero. 
Therefore, only when two branes are parallel along the third torus, the amplitude is non-zero. 

Therefore, from now on we consider the case where $v^3=0=v^{k;3}$. Then after spin-structure sum, 
as usual in $\N=2$ sectors 
the functions in the numerator cancel those in the denominator, 
and the entire $\nu$-dependence disappears from the integrand of the amplitude, for all worldsheets.  
Thus  the amplitudes reduce to the following:
\be \label{Z}
\langle \Phi_3 \bar \Phi_{\bar{3}} \rangle_{\A}^{\N=2} &=& \delta e_3 e_{\bar{3}} L_3^{-2}
\xi^{\A} \int_{1/\Lambda^2}^{\infty}\frac{dt}{t^2}\int_0^{t/2} d\nu \, \Gamma_\A(t,T^3,V_a^3) e^{-2\pi \chi t} \\
&=&{1 \over 2} \delta e_3 e_{\bar{3}} L_3^{-2} \xi^{\A} \int_{1/\Lambda^2}^{\infty}\frac{dt}{t} \, \Gamma_\A(t,T^3,V_a^3) e^{-2\pi \chi t}
\ee  
for annulus,
 and 
\be \label{L}
\langle \Phi_3 \bar \Phi_{\bar{3}} \rangle_{\M}^{\N=2} &=& - \delta e_3 e_{\bar{3}} L_3^{-2}
\xi^{\M} \int_{1/4\Lambda^2}^{\infty}\frac{dt}{t^2}\int_0^{t} d\nu \, \Gamma_\M(t,T^3,V_{O_k}^3) e^{-2\pi \chi t}  \\
&=&
- \delta e_3 e_{\bar{3}} L_3^{-2} \xi^{\M} \int_{1/4\Lambda^2}^{\infty}\frac{dt}{t}\, \Gamma_\M(t,T^3,V_{O_k}^3) e^{-2\pi \chi t}
\ee  
for M\"obius,  where the lattice sums $\Gamma_{\A}$ and $\Gamma_{\M}$ are given in \eqref{gammadefs} in the appendix.
Here the normalization constants $\xi$ are
\be
\xi^{\A} = - {g_o^2 \alpha' \over 8 N (4 \pi^2 \alpha')^2} c_{\A}  \; , \quad   \xi^{\M} =  - {g_o^2 \alpha' \over 8N (4 \pi^2 \alpha')^2} c_{\M}\ ,
\ee 
where $c_{\A}$ and $c_{\M}$ are the usual traces involving also the intersection numbers along the two tori with non-trivial angles (these terms become the beta functions for gauge fields). 

The calculation of (\ref{Z}) and (\ref{L}) was performed  
for example in \cite{Lust:2003ky} (section \ 3.3)
but since the angles do not play a role, we can also use results from  
\cite{Berg:2004ek}  that are summarized in the appendix:
\be
\langle \Phi_3 \bar \Phi_{\bar{3}} \rangle_{\A}^{\N=2} &=& 
- \delta e_3 e_{\bar{3}} {\xi^{\A} \over 2} \ln\left( {T_2^3 V_a^3 |\eta(T^3)|^4}\right) L_3^{-2}
\ee
for annulus\footnote{Comparing
to the explicit expression in the appendix, we have used a scheme for $t\rightarrow 0$ divergences
where we subtract $\ln(8\pi^3 \chi)$, where $\chi$ is the cutoff
in \eqref{Z} and \eqref{L}. This does not affect the moduli dependence, of course.
Also we dropped the $\Lambda$-terms that cancel by tadpole cancellation.} (note that the superscript $3$ stands for the third torus and not for the third power),
\be
\langle \Phi_3 \bar \Phi_{\bar{3}} \rangle_{\M}^{\N=2} &=& 
\delta e_3 e_{\bar{3}} {\xi^{\M} \over 4} \ln\left( {T_2^3 V_{O_k}^3|\eta(T^3)|^4}\right) L_3^{-2}
\ee
for M\"obius, 
and the total is 
\be
\langle \Phi_3 \bar \Phi_{\bar{3}} \rangle_{\A}^{\N=2} + 
\langle \Phi_3 \bar \Phi_{\bar{3}} \rangle_{\M}^{\N=2}\ .
\ee
This is of course consistent with the expectation \Ref{expect}, if we take into account the $T^3_2$-dependence of $L_3$, given in \eqref{LD}.
 
 
\section{Conclusions and Outlook}

In this paper we have computed $\N=1$ and $\N=2$ contributions to the
one-loop renormalization of the K\"ahler metric of D-brane moduli, and
shown that the $\N=1$ contributions vanish. The $\N=2$ contributions, that
exist for parallel branes only, do not vanish, but are given by some
explicit expressions in the K\"ahler moduli.

That these $\N=2$ contributions are present is no surprise, but the
vanishing of the $\N=1$ contributions appears nontrivial to us. 
It may represent an interesting statement about the underlying string
theory rather than a nonrenormalization theorem of the effective field
theory. Such statements are somewhat rare in string effective actions.

We do not know any symmetry arguments that the $\N=1$ contributions should vanish,
but it is possible that charge selection rules prohibit couplings
of the kind needed to generate these loop-level contributions.\footnote{We thank M.\ Goodsell 
for very helpful email discussions on this topic.}

In future work, it would be interesting  to also compute the analogous quantities
with magnetized branes instead of branes at angles. We see no clear reason that
the former should vanish, as the configurations are not T-dual in these nontrivial backgrounds.

In more general terms, it would be interesting to understand how robust this result is. One obvious test to subject it to would be to deform away from the orientifold point by adding infinitesimal blowup modes. Another direction would be to attempt the calculation at higher genus. 

The first obvious application is to D-brane inflation. One could a priori have worried that an analogue of these corrections in smooth backgrounds would produce additional contributions to the eta problem (see \cite{Baumann:2009ni}). Of course, we have not shown that this generalizes to smooth backgrounds,
but there are similar partial vanishing results in smooth backgrounds (see the appendix of \cite{Baumann:2007ah}) and one could pursue that connection further.

If the two-loop contribution does not vanish, and at the moment we see no reason why it should, one could picture one interesting kind of application of the nonrenormalization result in this paper,
in orbifolds where there are no $\N=2$ subsectors. In  \cite{Kachru:2007xp,Berg:2010ha} and related work, flavor physics is studied in this context. This is very challenging in a top-down approach; even if one can arrange good flavor structure at tree-level (for an explicit example see \cite{Dolan:2011qu}), it is not obviously enough, since it would be ruined by generic quantum corrections at a level that is still inconsistent with experiment. A familiar example of how nontrivial this can be is the GIM (Glashow-Iliopoulos-Maiani) mechanism in the Standard Model, by which flavor-changing neutral currents are suppressed to effectively two-loop order.  Of course, we have not shown that our result generalizes to visible-sector matter fields, and it may not. 

For this and other reasons, it would be interesting to apply the same techniques to calculating one-loop corrections to the  K\"ahler metric of chiral matter fields. One example of this direction can be found in \cite{Benakli:2008ub}.

In general, we find it important to further develop the technology for 
calculating moduli-dependent string effective actions with minimal supersymmetry.
As emphasized for example in \cite{DiVecchia:2008tm}, there are still many 
fundamental issues for which techniques are lacking. 

\vskip 15mm


\noindent
{\Large {\bf Acknowledgments}}

\vskip 3mm

\noindent
We would like to thank Ralph Blumenhagen, Joe Conlon, Fawad Hassan, Gabi Honecker, Stefan Sj\"ors and Stephan Stieberger for valuable discussions and email correspondence,
and Mark Goodsell for very useful comments on an early draft. MH thanks the Oskar Klein Center, Stockholm University, and NORDITA for hospitality. JK thanks the KITPC, Beijing, for hospitality. This work is supported in part by the Excellence Cluster ``The Origin and the Structure of the Universe'' in Munich. The work of MH is supported by the German Research Foundation (DFG) within the Emmy-Noether-Program (grant number: HA 3448/3-1). JK received support from the German exchange office (DAAD). 
MB  thanks the Swedish Research Council (VR) for support. We also gratefully acknowledge support for this work by the Swedish Foundation for International Cooperation in Research and Higher Education.

 
\appendix
 
\section{Variables from Reduction}
\label{app:reduction}
\subsection{K\"ahler variables}
In this appendix we perform a dimensional reduction of the DBI action to 
see what the natural variables are to work with. 
For the following we refer the reader to equations \eqref{AIIB} and  \eqref{L4} in the main text. 

We are interested in expanding $\sqrt{\det (P[G] + {\cal F})}$ to second order in the fluctuations along or transverse to the branes. Here ${\cal F} = F + P(B)$ and $P[G]$ and $P(B)$ stand for the pullbacks of the metric and $B$-field. Thus, we need
\be \label{PGtips}
P[G]_{\mu \nu} &=& G_{\mu \nu} + G_{a (\mu} \partial_{\nu)} \phi^a + G_{ab} \partial_{\mu} \phi^a \partial_{\nu} \phi^b = G_{\mu \nu} + G_{ab} \partial_{\mu} \phi^a \partial_{\nu} \phi^b\ , \nonumber \\
P[G]_{\mu A} &=& G_{\mu A} + G_{a (A} \partial_{\mu)} \phi^a + G_{ab} \partial_{A} \phi^a \partial_{\mu} \phi^b = G_{a A} \partial_{\mu} \phi^a\ , \\
P[G]_{AB} &=& G_{AB} + G_{a (A} \partial_{B)} \phi^a + G_{ab} \partial_{A} \phi^a \partial_{B} \phi^b = G_{AB} \nonumber \ .
\ee
Here we assumed that all fields only vary with respect to the external coordinates $X^\mu$ and not with respect to $X^A$. Moreover, the metric is supposed to have no off-diagonal entries with one external and one internal index. 

Moreover, we will need the components of the gauge field:
\be
{\cal F}_{\mu A} & = & F_{\mu A} + B_{\mu A} - B_{a[\mu} \partial_{A]} \phi^a + B_{ab} \partial_{\mu} \phi^a \partial_{A} \phi^b = \partial_\mu A_A- B_{Aa}  \partial_\mu \phi^a\ , \\
{\cal F}_{AB} & = & F_{AB} + B_{AB} - B_{a[A} \partial_{B]} \phi^a + B_{ab} \partial_{A} \phi^a \partial_{B} \phi^b = 0
\ee
and ${\cal F}_{\mu \nu}$. Here we used again that all fields only depend on $X^\mu$ and also the fact that we only consider untwisted components of the $B$-field.\footnote{That is, we expand the $B$-field only along the untwisted $(1,1)$-forms and not the twisted ones, cf.\ formula (2.1) in \cite{Gmeiner:2007zz}.} The untwisted $B$-field factorizes, i.e.\ there is one component along each torus. This implies that the only non-vanishing components are of the form $B_{aA}$, because $B_{AB}$ or $B_{ab}$ would have legs along two different tori.

\begin{figure}[th]
\begin{center}
\includegraphics[scale=0.75]{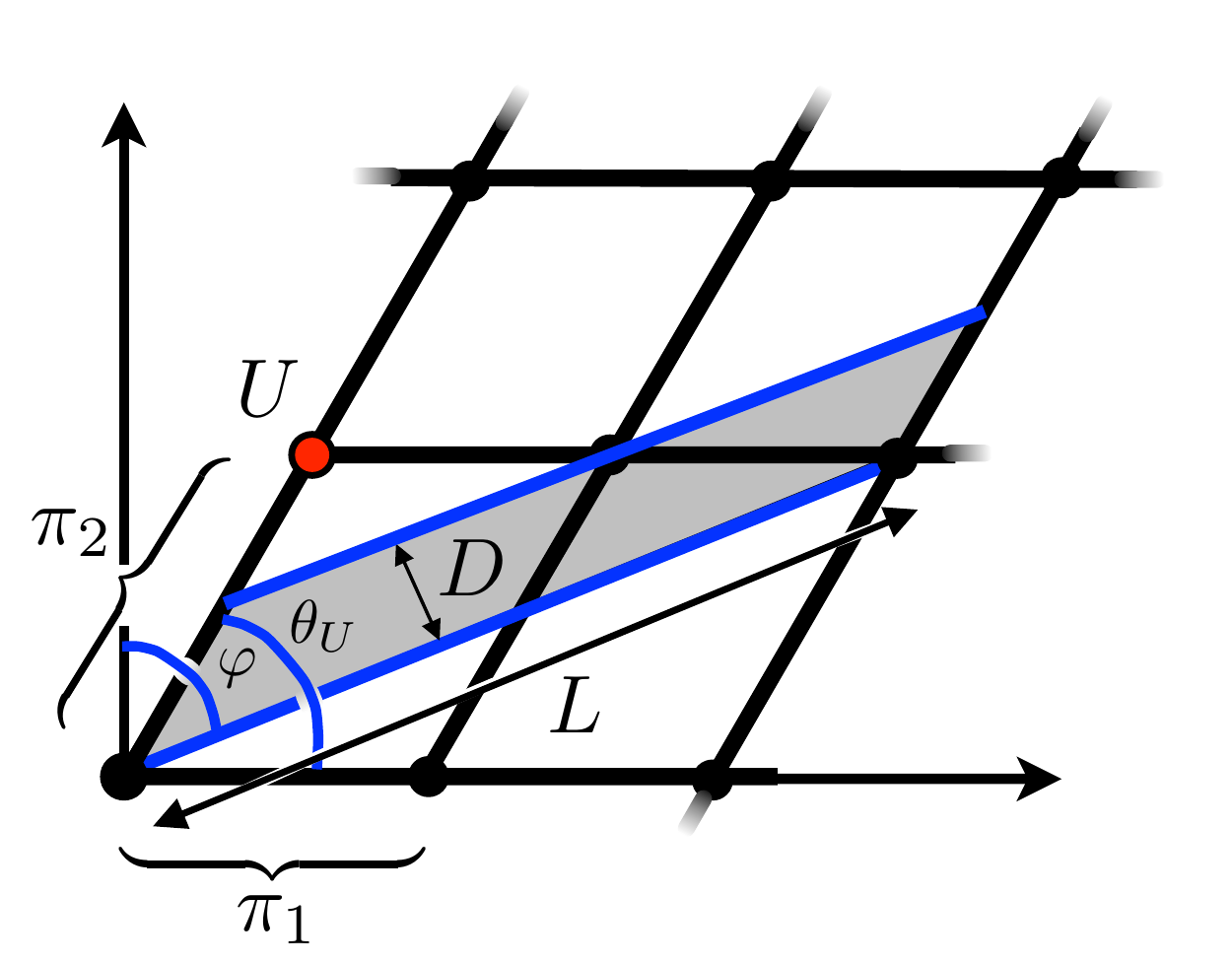}
\caption{\label{tiltedtorus} The wrapped brane with $(n,m)=(2,1)$.}
\end{center}
\end{figure}

In order to proceed further, we need the form of the torus metric. There are two metrics that are commonly used, cf.\ the discussion in chapter 5.1 of \cite{Polchinski:1998rq}. The first choice is
\be \label{t2metric}
ds^2 = \frac{\rho}{U_2} |d\tilde x^1 + U d \tilde x^2|^2 = \frac{\rho}{U_2} ((d \tilde x^1)^2 + 2 U_1 d \tilde x_1 d \tilde x_2 + |U|^2 (d \tilde x^2)^2) \ ,
\ee
where $\tilde x^1$ and $\tilde x^2$ are periodic with period $1$, for instance, and $\rho$ denotes the volume of the torus. Alternatively, the second choice is
\be \label{t2metric2}
ds^2 = \frac{\rho}{U_2} |dx^1 + i  d x^2|^2 = \frac{\rho}{U_2} ((dx^1)^2 + (dx^2)^2)\ ,
\ee
where now the periodicity of $x^1$ and $x^2$ is
\be \label{periodicity}
(x^1 + i x^2) \equiv (x^1 + i x^2) + (m + n U)\ .
\ee
In \eqref{PGtips}, we need the metric components using coordinates $X^A, X^a$ which are adapted to the worldvolume of the brane, i.e.\ coordinates along and transverse to the brane (in static gauge). Thus, it is more convenient to consider a different elementary cell for the torus, i.e.\ the region between two neighboring parts of the brane, cf.\ the shaded region in fig.\ \ref{tiltedtorus}. The corresponding metric is most conveniently chosen to be flat and diagonal, similar to the one in \eqref{t2metric2}. However, now the new elementary cell has length and hight  \cite{Blumenhagen:2006ci}
\be \label{LD}
L^2 = \frac{\rho}{U_2} |n+m U|^2 \quad , \quad D^2 = \frac{\rho^2}{L^2} = \frac{\rho U_2}{|n+m U|^2}\ .
\ee
The integers
$n$ and $m$ are the wrapping numbers of the brane under consideration which wraps the cycle
\be
\Pi = n \pi_1 + m \pi_2\ .
\ee
From fig.\ \ref{tiltedtorus}, we see that the complex structure for the new elementary cell is 
\be
\tilde U \equiv \frac{D}{L} \left( \frac{1}{\tan(\theta_U-(\pi/2-\varphi))} + i \right) = \frac{D}{L} \left( \frac{1}{-\cot(\theta_U+\varphi)} + i \right)\ .
\ee
Comparing with \eqref{t2metric2} and noting that 
\be
\frac{\rho}{\tilde U_2} = \frac{DL}{D/L} = L^2\ ,
\ee
one might be tempted to use the metric 
\be \label{metrictempt}
ds^2 = L^2 ((dx^A)^2 + (dx^a)^2)\ ,
\ee
with periodicity 
\be \label{periodicity2}
(x^A + i x^a) \equiv (x^A + i x^a) + (m + n \tilde U)\ .
\ee
However, this definition would have the disadvantage that some of the moduli dependence of the low energy effective action would be hidden in the integration region of $x^a$. To avoid this, one should rescale the coordinate $x^a$ in such a way that $x^a=1$ corresponds to the {\it physical} length $D$, in the same way as $x^A=1$ corresponds to the length $L$. Comparing with \eqref{periodicity2} we see that this can be done by rescaling $x^a$ by $1/\tilde{U}_2=L/D$. This, on the other hand, implies a change of the metric which becomes
\be \label{t2metric3}
ds^2 = L^2 (dx^A)^2 + D^2 (dx^a)^2\ .
\ee
 
Now we have for the DBI action
\be
&& \hspace{-0.5cm} \sqrt{\det\left[ \left( \begin{array}{cc}
                                 G_{\mu \nu}  & 0 \\
                                 0 & G_{AB}
                                 \end{array}
                       \right) 
                       + \left(  \begin{array}{cc}
                                    {\cal F}_{\mu \nu}+ G_{ab} \partial_{\mu} \phi^a \partial_{\nu} \phi^b & G_{Ba} \partial_{\mu} \phi^a + (\partial_\mu A_B- B_{Ba}  \partial_\mu \phi^a) \\
                                    G_{Aa} \partial_{\nu} \phi^a - (\partial_\nu A_A- B_{Aa}  \partial_\nu \phi^a) & 0
                                    \end{array}
                       \right)\right]} \nonumber \\
&=& \sqrt{\det G_{\rho \sigma}}  \sqrt{\det G_{CD}} \Big( 1 - \tfrac14 {\cal F}_{\mu \nu} {\cal F}^{\mu \nu} + \tfrac12 G_{ab} \partial_{\mu} \phi^a \partial^{\mu} \phi^b \nonumber \\
&& \hspace{2cm} +\tfrac12 \Big( (\partial_\mu A_B - B_{Ba}  \partial_\mu \phi^a)(\partial_\nu A_A - B_{Aa}  \partial_\nu \phi^a) - G_{Aa} G_{Bb} \partial_{\nu} \phi^a \partial_{\mu} \phi^b \Big) G^{\mu \nu} G^{AB} \Big)\label{kinetic} \ .
\ee

The first term in \eqref{kinetic} is a contribution to the potential which is cancelled once tadpole cancellation is imposed. We now note that $G_{a A}=0$ and  
\be
G_{ab} \sim \delta_{ab} \quad , \quad G_{AB} \sim \delta_{AB} \ ,
\ee
\be
G_{ab} \sim \left\{\begin{array}{ll} \neq 0 & a=b \\ = 0 & a\neq
b\end{array} \right.
 \; , \quad 
 G_{AB} \sim \left\{\begin{array}{ll} \neq 0 & A=B \\ = 0 & A\neq
B\end{array} \right.
 \quad    \ ,
\ee
as the 6-torus is a product of three 2-tori. Thus, denoting the Wilson line along the $\alphai$th brane with $A_\alphai$ and the position modulus with $\phi^\alphai$, the kinetic terms of the scalars can be rewritten as
\beqn
&& \sqrt{\det G_{\rho \sigma}}  \sqrt{\det G_{CD}} \sum_{\alphai} \Big( \tfrac12 G^\alphai_{22} \partial_{\mu} \phi^\alphai \partial^{\mu} \phi^\alphai \nonumber \\
&&  \hspace{3.5cm} +\tfrac12 \Big( (\partial_\mu A_i - B_\alphai \partial_\mu \phi^\alphai)(\partial_\nu A_i - B_\alphai \partial_\nu \phi^\alphai) \Big) G^{\mu \nu} \frac{1}{G^\alphai_{11}} \Big) \nonumber
\eeqn
This can be simplified using \eqref{LD} and \eqref{t2metric3}, resulting in 
\beqn
&& \sqrt{\det G_{\rho \sigma}}  \sqrt{\det G_{CD}} \sum_{\alphai} \tfrac12 \Big( (\partial_\mu A_i - B_\alphai \partial_\mu \phi^i)(\partial^\mu A_i - B_\alphai \partial^\mu \phi^i) +\rho_\alphai^2 \partial_{\mu} \phi^i \partial^{\mu} \phi^i  \Big) \frac{1}{L^2_{\alphai}}\nonumber \\
&=& \sqrt{\det G_{\rho \sigma}}  \sqrt{\det G_{CD}} \sum_{\alphai} \frac{1}{2 L^2_{\alphai}} |T_\alphai \partial \phi^i - \partial A_i|^2 \label{phiAkinetic}
\eeqn
with 
\be \label{Tdef}
T_\alphai = B_\alphai + i \rho_\alphai\ ,
\ee
where $B_\alphai$ denotes the component of the B-field along the $\alphai$th torus,
and $\rho_{\alphai}$ is the volume of the $\alphai$th torus. The result \eqref{phiAkinetic} is used in sec.\ \ref{Kvariables} in order to argue for the form of the variables \eqref{AIIA}.

In order to obtain the kinetic term in the Einstein frame, one still has to perform a Weyl rescaling. Although
we will not need this in detail, let us end this appendix
by a closer look at this rescaling. The kinetic terms of the vectors and scalars are
\be
&& \hspace{-1.5cm} \int d^4 x\, \sqrt{\det G_{\rho \sigma}} \Big( e^{-\Phi} \int_\Sigma d^3 \xi \, \sqrt{\det G_{CD}} \Big) \Big[ - \tfrac14 {\cal F}_{\mu \nu} {\cal F}^{\mu \nu} + \sum_{\alphai}  \frac{1}{2 L^2_\alphai} |T_\alphai \partial \phi^i - \partial A_i|^2 \Big] \ ,
\ee
where $\Sigma$ is the cycle wrapped by the brane stack. The gauge coupling is given by the volume of the 3-cycle wrapped by the brane stack. For a brane wrapping a calibrated 3-cycle, this volume can also be expressed as \cite{Blumenhagen:2006ci}
\be \label{3cycle}
e^{-\Phi} \int_\Sigma d^3 \xi \, \sqrt{\det G_{CD}} = e^{-\Phi} \int_\Sigma {\rm Re}\,  \Omega = e^{-\Phi} \sqrt{\prod_\alphai \rho_\alphai U_{2 \alphai}^{-1} |n_\alphai + U_\alphai m_\alphai|^2 }=e^{-\Phi} \prod_\alphai L_\alphai\ ,
\ee
which depends on the complex structure of the Calabi-Yau orientifold. After a Weyl rescaling the kinetic term for the scalars takes the form
\be
&& \hspace{-1.5cm} \int d^4 x\, \sqrt{\det G_{\mu \nu}} \  \sum_{{\rm branes}} \frac{\int_{\Sigma} {\rm Re}\, \Omega}{2 e^{-\Phi} {\cal V}} \sum_i \frac{|T_i \partial \phi^{i} - \partial A_i|^2}{L^2_\alphai} + \ldots \ .
\ee

The prefactor $\frac{\int_{\Sigma} {\rm Re}\, \Omega}{e^{-\Phi} {\cal V}}$ scales like the inverse of a 3-cycle volume and it should be possible to express it in terms of the complex structure of the orientifold. Note that writing it as
\be
\frac{e^{\Phi} \int_\Sigma d^3 \xi \, \sqrt{\det G_{AB}}}{\int_Y d^6 X \, \sqrt{\det G_{AB} \det G_{ab}}}\ ,
\ee
it is analogous to the prefactor in (4.85) of \cite{Blumenhagen:2006ci}.


\section{Variables}
\label{variables:appendix}

It is important conceptually that one could work with vertex operators that are adapted to 
the intrinsic coordinates of the brane
and make no reference to the ambient space. 
However, there is a certain tension between this and using complex embedding coordinates
where target space rotations are simple. For our calculation
we have used the latter, but for completeness, here we discuss how
we can use the former. 
We follow \cite{Hassan:2003uq} and, for ease of notation, we only consider branes in a non-compact spacetime so that we do not have to distinguish between coordinate $X^\mu$ and $X^A$ as in \eqref{coordinates}.
Let the spacetime coordinates be denoted by $X^M$, the D-brane worldvolume coordinates by $\zeta^{\alpha}$,
and D-brane embedding function by $X^M(\zeta^{\alpha})$. We can introduce a set of normal vectors $a^M_I$
that are orthogonal to  $\partial_{\alpha} X^M$ in the spacetime metric $G_{MN}$, and normalize them:
\be
\partial_{\alpha} X^M G_{MN} a^N_I = 0 \; ,  \quad
a^M_I G_{MN} a^N_I = \delta_{IJ} \;.   \label{defa}
\ee
We can write intrinsic field variables $\hat{A}_{\alpha}$, $\hat{\phi}^I$
\be
A_M a^M_I = 0 \; , \quad A_M \partial_{\alpha}X^M = \hat{A}_{\alpha} \; , \quad \phi^M a_M^I = \hat{\phi}^I\ .
\ee
In static gauge, $X^{\mu} = \zeta^{\mu}$, then $A_{\mu}=\hat{A}_{\mu}$. 
We convert  back by 
\be
\hat{A}_{\alpha} g^{\alpha \beta} \partial_{\beta} X^M G_{MN} = A_N
 \; , \quad
 \hat{\phi}^I a_I^M = \phi^M
\ee
where $g_{\alpha \beta}$ is the induced metric on the worldvolume.
Amongst other things, we note that $\phi^{\mu} \neq 0$, but $\phi_{\mu}=0$.\footnote{Note that  standard Buscher rules mix closed string fields $G$ and $B$
under T-duality, but they do not mix in 
the open-string field $F$. 
One can still use them with $F$ as follows. Set $F=0$, $B\neq0$, perform T-duality,
then perform an $O(d,d)$ gauge transformation to map $B\rightarrow F$.  Here we 
will (as a first attempt) not use the Buscher rules at all but rather act directly on the worldsheet fields,
which is how the Buscher rules are derived in the first place.}
 
The boundary conditions are\footnote{Here $\sigma^{\pm} = {1 \over 2}(\tau\pm \sigma)$,
so in (42) in \cite{Hassan:2003uq}
 , $\partial_{-}X-\partial_{+}X = -\partial_{\sigma}X$.}
 $\partial_{\alpha}X^LN_L = 0$ and $a_L^I D^L=0$, where
\be
N_L &=& G_{LN}\partial_{\sigma} X^N + F_{LN} \partial_{\tau}X^N \qquad \mbox{(on boundary)}\ , \\
 D^L &=& \partial_{\tau}X^L \qquad \mbox{(on boundary)}\ ,
\ee
whereas the projections that survive are
\be
N_I &=& a^L_I N_L  \qquad \qquad\qquad \mbox{(normal)}\ , \\
D^{\alpha} &=& g^{\alpha \beta}\partial_{\beta}X^M G_{ML}D^L
\qquad  \mbox{(parallel)} \ ,
\ee
and with these, the boundary
couplings are
\be
N_I \hat{\phi}^I \; , \quad  D^{\alpha}\hat{A}_{\alpha} \;. 
\ee

We expect vertex operators for the intrinsic coordinates to then be converted as above:
\be
V_{\hat{A}^{\alpha}} &=& G_{MN} e_{\alpha }^N V_{A^M} \\
V_{\hat{\phi}^I} &=& a^I_M V_{\phi^M}
\ee
where ${e_{\alpha}}^N = \partial_{\alpha} X^N$. 
As stated before,  we will work with the vertex operators \eqref{Vops} in the ambient space. 


\section{Tadpole cancellation in $\mathbb{Z}_6'$ Orientifold}
\label{app:tadpole}
\subsection{Setup}

\begin{figure}
\includegraphics[width=0.9\textwidth]{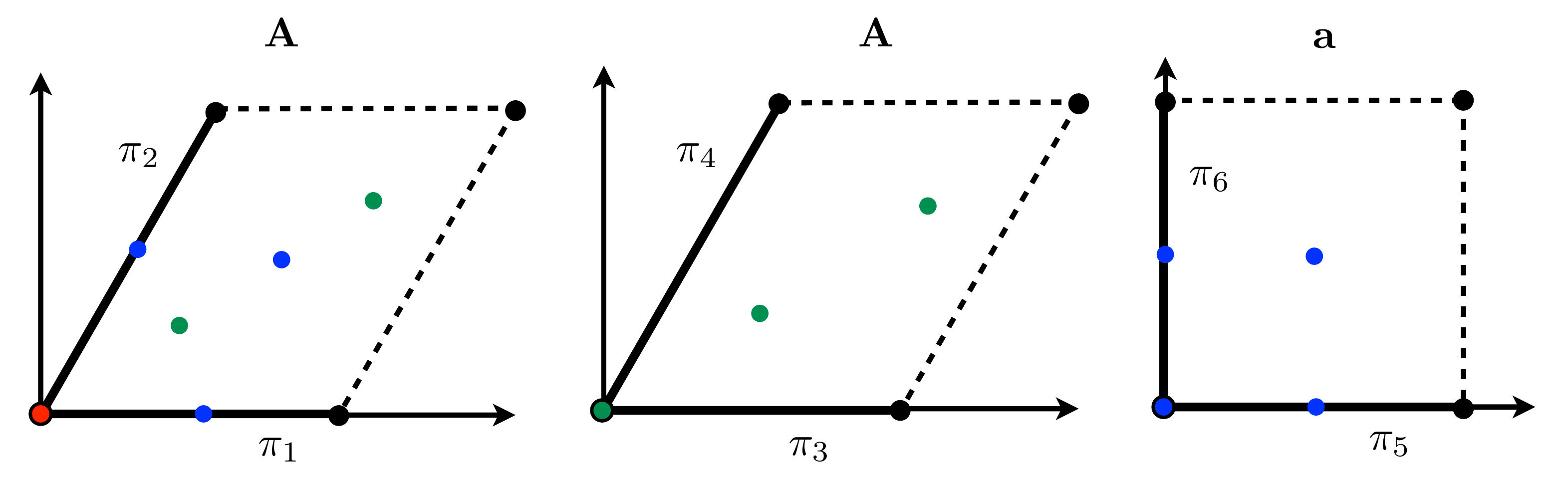}
\caption{
The AAa lattice.}
\label{fig:latt}
\end{figure}
In this section we give some background on the $T^6/\mathbb{Z}_6'$ orientifold, following \cite{Gmeiner:2007zz} closely. We take the orbifold generator of the $\mathbb{Z}_6'$ orientifold to be defined via the vector
\be
\vec{v} = \left( \frac16, \frac13, -\frac12 \right)\ .
\ee
There are a few different implementations
of the $\mathbb{Z}_6'$ orientifold, we will focus
on what is known as the AAa lattice (see for example
\cite{Gmeiner:2007zz} for more on the classification).
We show this lattice with orbifold fixed points
in figure \ref{fig:latt}.

We will consider branes wrapped only on {\it bulk cycles}, that
are not shrunk to zero, as opposed to
{\it fractional cycles}. The bulk cycles are inherited from the covering torus
of the orbifold. 
For $T^6/\mathbb{Z}_6'$, 
the invariant forms are $dz_1 \wedge dz_2 \wedge dz_3$,
$dz_1 \wedge dz_2 \wedge d\bar{z}_3$
and their complex conjugates (see e.g. \cite{Bailin:2006zf}), so
we have $b^{3,0}=b^{0,3}=b^{2,1}=b^{1,2}=1$, i.e.\ a total of 
four three-cycles.
We can expand a generic three cycle (not necessarily invariant) either in terms
of the six elementary 1-cycles of the covering six-torus
\be \label{Pibulk}
n_1 \pi_{1,0,0} + m_1 \pi_{2,0,0}+
n_2 \pi_{0,3,0} + m_2 \pi_{0,4,0}+
n_3 \pi_{0,0,5} + m_3 \pi_{0,0,6}\ ,
\ee
with the {\it wrapping numbers} $(n_i,m_i)$,
or in terms of the four basis cycles $\rho_i$ as
\be \label{PQ}
\Pi^{\rm bulk}=P\rho_1+Q \rho_2 + U \rho_3 + V \rho_4
\ee
for some integer expansion coefficients $(P,Q,U,V)$.\footnote{In order to comply with the notation in \cite{Gmeiner:2007zz} we denote the basis cycles by $\rho$, even though we also use $\rho$ for the volume moduli of the tori, cf.\ \eqref{Tmodulus} and for the Chan-Paton phase in the M\"obius amplitude, cf.\ \eqref{CPphase}. As the basis cycles only appear in this appendix, we feel that it should be always clear from the context what we mean.}
For the basis cycles $\rho_i$ to be invariant under the orbifold action $\twist$
we can form {\it orbits} by acting on the elementary cycles as
\be
\rho_1 &=& (1+\twist+ \twist^2 + \twist^3 + \twist^4 + \twist^5) \pi_{1,3,5} \\
&=& 2(1+\twist+ \twist^2)\pi_{1,3,5}\\
&=& 2 \pi_{1,3,5}+2  \pi_{2,4-3,5} +2  \pi_{2-1,-4,5}\ .
\ee
In total the four orbits are
\be
\rho_1 &=& 2 \pi_{1,3,5}+2  \pi_{2,4-3,-5} +2  \pi_{2-1,-4,5}\ , \\
\rho_2 &=& 2\pi_{1,4,5}+2  \pi_{2,-3,-5} +2  \pi_{2-1,3-4,5}\ , \\
\rho_3 &=& 2 \pi_{1,3,6}+2 \pi_{2,4-3,-6} +2  \pi_{2-1,-4,6}\ , \\
\rho_4 &=& 2 \pi_{1,4,6}+2  \pi_{2,-3,-6} +2  \pi_{2-1,3-4,6}\ .
\ee
At this point it is clear why there are {\it three} members $a^k$ of each orbit $[a]$, 
i.e.\ three terms in each line above ---
 it corresponds to the action $\twist^k$ for $k=0,1,2$. 
These cycles can also be decomposed in terms of the elementary cycles as
\be
\rho_1 &=& 2 \pi_{1,3,5}-4  \pi_{2,4,5} +2  \pi_{2,3,5} +2  \pi_{1,4,5}\ , \\
\rho_2 &=& 4 \pi_{1,4,5}+4  \pi_{2,3,5} -2  \pi_{2,4,5} -2  \pi_{1,3,5}\ , \\
\rho_3 &=& 2 \pi_{1,3,6}-4  \pi_{2,4,6} +2  \pi_{2,3,6} +2  \pi_{1,4,6}\ , \\
\rho_4 &=& 4 \pi_{1,4,6}+4  \pi_{2,3,6} -2  \pi_{2,4,6} -2  \pi_{1,3,6}\ , 
\ee
so we can sum over either. This latter representation is  perhaps less intuitive
(there are now {\it four} terms in each basis cycle), but convenient: we can easily 
see which cycles intersect. To do so, recall that self-intersection is zero,
so a nonvanishing example is $\pi_{1,3,5} \circ \pi_{2,4,6}=(\pi_1 \circ \pi_2)(\pi_3 \circ \pi_4)(\pi_5 \circ \pi_6)=
1$. Our conventions are \ $\pi_1 \circ \pi_2 = -\pi_2 \circ \pi_1 = 1$ and so on. 
We then easily establish that
\be
\rho_1 \circ \rho_3 &=&(-1)^3 {1 \over 6} \Big[(2 \pi_{1,3,5}) \circ (-4\pi_{2,4,6})+ ( -4\pi_{2,4,5}) \circ (2 \pi_{1,3,6}) \\
&&\qquad \quad +(2\pi_{2,3,5}) \circ (2\pi_{1,4,6}) +(2 \pi_{1,4,5})\circ (2\pi_{2,3,6})\Big] \\
&=&(-1)^3 {1 \over 6} [-8\cdot 1-8\cdot 1+4\cdot (-1)+4\cdot(-1)]=4 \; . 
\ee
Continuing like this, the intersection matrix becomes
\be
I_{\rho_i \rho_j} = \rho_i \circ \rho_j =
\begin{pmatrix}  
0 & 0 & 4 & 2  \\ 
0 & 0 & 2 & 4  \\ 
-4 & -2 & 0 & 0  \\ 
-2 & -4 & 0 & 0  \\ 
 \end{pmatrix} \; . 
\ee
Comparing \Ref{Pibulk} and \Ref{PQ} we  can relate the expansion coefficients $(P,Q,U,V)$ and the wrapping numbers.
To do so, it is convenient to note that the action of $\Theta$ on the wrapping numbers is (cf.\ eq.\ (2.2) in \cite{Gmeiner:2007zz}).
\be
\label{wrapZ6}
\begin{pmatrix}
n_1 & m_1 \\
n_2 & m_2 \\
n_3 & m_3 
\end{pmatrix}
\, , \; k=1:
\begin{pmatrix}
-m_1 & n_1+m_1 \\
-(n_2+m_2) & n_2 \\
-n_3 & -m_3 
\end{pmatrix}
\, , \; k=2:
\begin{pmatrix}
-(n_1+m_1) & n_1 \\
m_2 & -(n_2+m_2) \\
n_3 & m_3 
\end{pmatrix} .
\ee
We can then extract expressions for the expansion coefficients in \Ref{PQ}
in terms of wrapping numbers:
\be
P &=& (n_1 n_2-m_1 m_2)n_3\ , \\
Q &=& (n_1 m_2 + m_1 n_2 + m_1 m_2)n_3\ , \\
U &=& (n_1 n_2 - m_1 m_2) m_3\ , \\
V &=& (n_1 m_2 + m_1 n_2 +m_1 m_2)m_3\ .
\ee
As an example, the cycle with wrapping numbers
$(1,0;1,0;1,0)$ produces $P=1$ and $Q=U=V=0$, i.e\ it corresponds
to $\rho_1$. What this means
is that $(1,0;1,0;1,0)$ is one representative in the collection
of wrapping numbers that forms the orbit $\rho_1$. 

Using the intersection numbers for the $\rho$, we can compute
\be
I_{ab} = \Pi_a^{\rm bulk} \circ \Pi_b^{\rm bulk} &=& 
(P_a \rho_1+Q_a \rho_2 + U_a \rho_3 + V_a \rho_4) \circ
(P_b \rho_1+Q_b \rho_2 + U_b \rho_3 + V_b \rho_4) \nonumber \\
&=&
2(P_a  V_b +Q_a  U_b)
+4(P_a  U_b +Q_a  V_b) - (a \leftrightarrow b)\ .
\ee
These are our desired intersection numbers
of orbifold invariant collections of cycles. 

Finally, we want orbifold invariant collections of orientifold planes
and their wrapping numbers. 
The reflection $\R$ acts as $(\pi_1,\pi_2) \rightarrow (\pi_1,\pi_1-\pi_2)$ in
the first two 2-tori and as $(\pi_1,\pi_2) \rightarrow (\pi_1,-\pi_2)$ in the third.
This means for the wrapping numbers
\be
\label{wrapZ6R}
\begin{pmatrix}
n_1 & m_1 \\
n_2 & m_2 \\
n_3 & m_3 
\end{pmatrix}
\stackrel{\R}{\longrightarrow}
\begin{pmatrix}
n_1+m_1 & -m_1 \\
n_2+m_2 & -m_2 \\
n_3 & -m_3 \\
\end{pmatrix} .
\ee
The $\R$ images for the AAa lattice are then found as:
\be
\rho_1  &\rightarrow& \rho_1 \ , \\
\rho_2  &\rightarrow& \rho_1-\rho_2 \ , \\
\rho_3  &\rightarrow& -\rho_3 \ , \\
\rho_4  &\rightarrow& \rho_4-\rho_3  \ .
\ee
We want to form invariant combinations. Obviously
$\rho_1$ and $\rho_2$ transform among themselves,
and so do $\rho_3$ and $\rho_4$, so we expect 
two sub-orbits. The first one can obviously be chosen
to be $\rho_1$, and the second can be chosen as $\rho_3-2\rho_4$. 
The representatives of these two orbits are mapped into 
each other by even powers $\R \twist^{\rm even}$ and odd powers $\R \twist^{\rm odd}$,
respectively, so we label them by this. The wrapping numbers are:
\be
\Omega \R \twist^{\rm even}: \rho_1 && (n_i,m_i)= (1,0;1,0;1,0) \ , \\
\Omega \R \twist^{\rm odd}: \rho_3-2\rho_4 &&  (n_i,m_i)=  (1,1;0,1;0,-1)  \ . 
\ee
The representatives are generated from \eqref{wrapZ6}. The complete set of O6-plane wrapping numbers is:
\be  \label{O6expl}
\left[\begin{array}{rr} 1 & 0 \\ 1 & 0 \\ 1 & 0 \end{array} \right] \stackrel{\Theta}{\rightarrow}
\left[\begin{array}{rr} 0 & 1 \\ -1 & 1 \\ -1 & 0 \end{array} \right] \stackrel{\Theta}{\rightarrow}
\left[\begin{array}{rr} -1 & 1 \\ 0 & -1 \\ 1 & 0 \end{array} \right]\ , \\
\left[\begin{array}{rr} 1 & 1 \\ 0 & 1 \\ 0& -1 \end{array} \right] \stackrel{\Theta}{\rightarrow}
\left[\begin{array}{rr} -1 & 2 \\ -1 & 0 \\ 0 & 1 \end{array} \right] \stackrel{\Theta}{\rightarrow}
\left[\begin{array}{rr} -2 & 1 \\ 1 & -1 \\ 0 & -1 \end{array} \right]\ . 
\ee
It may also be useful to note that if we act further with $\Theta$, we can obtain wrapping
numbers that differ from these by an even number of sign flips, but as three-cycles, those are 
equivalent to this set of six. 


\subsection{Trigonometry}
The angle $\varphi$ between the brane $a$ and the $x$-axis is determined by 
\be
\cos \varphi = {nR_1 + mR_2 \cos \theta_U \over {\mathcal V}} \; , \quad
\sin \varphi = {mR_2 \sin \theta_U \over {\mathcal V}} \ ,
\ee
where $\theta_U$ is the angle that $U$ makes with the $x$-axis,
and ${\mathcal V}=R_1 L $ is the physical length of the cycle, i.e.\
{\it before} we scaled the coordinates so the horizontal basis vector is unit length
 ($\vec{e}_1 \rightarrow \vec{e}_1/|\vec{e}_1| = \vec{e}_1/R_1$). 
Now it is easy to compute
\be
\cot(\varphi_b-\varphi_a) &=& {\cos\varphi_b \cos \varphi_a+\sin\varphi_b \sin \varphi_a \over
\cos\varphi_a \sin \varphi_b - \cos \varphi_b \sin \varphi_a} \\[2mm]
&=&
{ n_a n_b{R_1 \over R_2} + m_a m_b {R_2 \over R_1} + (n_a m_b + n_b m_a)\cos \theta_U
\over I_{ab} \sin \theta_U}\\
&=&
{  V_{ab}
\over I_{ab} } \; ,  \label{cotv}
\ee
where we introduced
\be \label{Vab}
V_{ab}=
{1 \over \sin{\theta}} \left(
n_a n_b{R_1 \over R_2} + m_a m_b {R_2 \over R_1}+ (n_a m_b + n_b m_a)\cos \theta_U \right)  \; . 
\ee
For $a=b$, this is the {\it square} of the length of  brane $a$ which can alternatively be expressed as
\be \label{Vaa}
V_{aa} \equiv V_{a} = \frac{|n_a + U m_a|^2}{U_2}\ .
\ee
For $a\neq b$, it
has no particular meaning. 

Note the occurrence of the intersection numbers $I_{ab}=n_am_b-n_bm_a$; this comes from the `$\sin(\varphi_b-\varphi_a)$' in
the denominator. Also note the special cases\footnote{Correct with (12) in \cite{Gmeiner:2009fb}.}
\be
V_{ab} = \left\{ 
\begin{array}{ll}
n_a n_b{R_1 \over R_2} + m_a m_b {R_2 \over R_1} & (\theta_U=\pi/2) \\
{2 \over \sqrt{3}} \left(n_a n_b{} + m_a m_b {}  
+  {1 \over 2}(n_a m_b + n_b m_a)\right) \; . 
 & (\theta_U=\pi/3) 
 \end{array}
 \right.
\ee
where the last expression follows from $U\equiv R_2/R_1 e^{i\theta_U} = e^{\pi i /3}$ for $\mathbb{Z}_3$,
i.e.\ $R_2/R_1=1$. One can now write completely analogous
expressions for $V_{a,O_k}$, where $O_k$ is one of the orientifold planes \eqref{O6expl}. 


\subsection{Partition functions}
We consider the brane stack $a$ as a representative of the orientifold orbit $[a]$. 
The complete partition function is given in terms of vacuum amplitudes
on each worldsheet surface:
\be
\cz= \sum_{a\in [a], b \in [b]}\A_{a,b} + \sum_{a \in [a]}\sum_{k=0}^{5} \M^k_{a, \R a^k} + \K + \ct \;  .
\ee
The external partition functions are \cite{Antoniadis:1999ge}
\be \label{Z4}
\cz^{{\rm ext}}\zba{\alpha}{\beta}(\tau)=\frac{1}{(4 \pi^2 \alpha' t)^2} \frac{\thba{\alpha}{\beta}(0, \tau)}{\eta(\tau)^3}\ ,
\ee
where $\tau$ is $\tau_\A$ for annulus and $\tau_\M$ for M\"obius. 
The annulus partition function is
\be \label{Aab2}
\cz^{{\rm int}}_{\A}\zba{\alpha}{\beta} (\tau_\A)= \prod_{i=1}^3 I^i_{a,b}
\frac{\thba{\alpha+v^i_{ab}}{\beta}{(0,\tau_\A)}}{\thba{1/2+v^i_{ab}}{1/2}{(0,\tau_\A)}}\ ,
\ee 
where $\tau_\A=it/2$ and the angles are\footnote{Note that this definition differs by a factor of $i$ from the definition in \cite{Lust:2003ky}.}
\be \label{vab}
v^j_{ab} \equiv {1 \over \pi}(\varphi_a^j-\varphi_b^j) = {1 \over \pi} \varphi_{ab}^j \; . 
\ee
This contains the rotation angle $\varphi$ that depends on the representatives $a$ and $b$.

We note that when the supersymmetry condition $\sum_i^3 \varphi_{ab}^i=0$ mod $\pi$ holds,
we can rewrite this as  equation (2.13) in \cite{Lust:2003ky}.\footnote{In their eq.\ (2.17), we set $d=3$ for intersection in all three tori, $\epsilon\rightarrow 0$, $\beta\rightarrow 0$
as there is no external gauge field, 
and $Z_i=1$ as there are no zero modes.} 
There, the angle appears in the argument instead of the upper characteristic.

Now we need the annulus partition function
for strings stretching between some brane $a$ and its image $a'$. This could be either
the orbifold image $a^k$,
or the orientifold image thereof, $\Omega' \twist^k a=: \R a^k$. We need only specialize the expression (\ref{Aab2}) to
this case:
\be
\label{Aaa2}
\cz^{{\rm int}}_{\A}\zba{\alpha}{\beta} (\tau_\A)= \prod_{i=1}^3 I^i_{aa'}
\frac{\thba{\alpha+v^i_{aa'}}{\beta}{(0,\tau_\A)}}{\thba{1/2+v^i_{aa'}}{1/2}{(0,\tau_\A)}}\ .
\ee 
Even when $a$ is rotated by an angle $\varphi$ relative
to {\it another} system of branes (or O-planes),
this of course does not affect the angle between a representative
$a$ and its image $a'$, so
the $\A_{a,a'}$ amplitude does not depend on the rotation angle $\varphi$ directly. 

For the M\"obius vacuum amplitude\footnote{In their eq.\ (2.23), since $d'=0$, there is no product over $i$,
so no dependence on $n_{O6}$.}
for strings stretching from brane $a$ to the orientifold image $\Omega \R \twist^k a$ we have
\be \label{Z_int_M}
\cz_{\M}^{{\rm int},k}\zba{\alpha}{\beta} (\tau_\M)= \prod_{j=1}^3 I_{a, O}^{k;j}
\frac{\thba{\alpha+2 v^j_{a, O_k}}{\beta-v^j_{a, O_k}}{(0,\tau_\M)}}{\thba{1/2+ 2 v^j_{a, O_k}}{1/2-v^j_{a, O_k}}{(0,\tau_\M)}}\ ,
\ee
where
\be \label{vOab}
v^j_{a O_k} = -{1 \over \pi}(\varphi_a^j-\varphi_{O_k}^j)
\ee
and $\tau_\M=it/2+1/2$. Note that unlike \eqref{vab}, this explicitly depends on the sector $k$. 
As before, the internal part
can be rewritten with the shift in the argument instead of in the characteristics
provided $\sum_i v^i = 0$, and then it can be checked with \cite{Lust:2003ky}. 
Here $I_{a, O}^{k;j}$ is the number of 
$\Omega \cR \twist^k$-invariant intersections of the two branes. 

If along the $i$th torus the intersecting angle vanishes, the internal partition functions
(\ref{Aab2}) and (\ref{Z_int_M}) are modified as follows:
\be   \label{Amod}
\frac{I^i_{ab}}{\thba{1/2+v^i_{ab}}{1/2}{(0,\tau_\A)}} \stackrel{v^i_{ab}=0}{\Longrightarrow} \frac{\Gamma_\A(t,T^i,V_a^i)}{\eta^3(\tau_\A)}
\ee
for annulus and 
\be    \label{Mmod}
\frac{I_{a, O}^{k;i}}{\thba{1/2+2 v^{i}_{a O_k}}{1/2-v^{i}_{a O_k}}{(0,\tau_\M)}} \stackrel{v^{i}_{a O_k}=0}{\Longrightarrow} \frac{\Gamma_\M(t,T^i,V_{O_k}^i)}{\eta^3(\tau_\M)}
\ee
for M\"obius. The zero mode contributions
 $\Gamma_\A$ and $\Gamma_\M$ to the partition functions are\footnote{see for example (2.19) and (2.27) in 
\cite{Lust:2003ky}}
 \be  \label{gammadefs}
\Gamma_\A(t,T^i,V^i_a) &=& \sum_{m,n} e^{-{\pi t \over T_2^i V_a^i} |m+T^i n|^2}
=  \sum_{\vec{n}}e^{-\pi t  \vec{n}^{\rm T}G_{\A}\vec{n} }  = \tht(itG_{\A}) \\
\Gamma_\M(t,T^i,V_{O_k}^i) &=&  \sum_{m,n} e^{-{\pi t \over T_2^i V_{O_k}^i} |m+T^i n|^2}
=  \sum_{\vec{n}}e^{-\pi t  \vec{n}^{\rm T}G_{\M}\vec{n} } = \tht(itG_{\M})
\ee
where in the latter sum, $V_{O_k}^i$ refers to O-planes that are parallel to D-branes 
(in these $\N=2$ sectors), and 
\be
G_{\A} = {1 \over T_2 V_a^i} \left( \begin{array}{cc} 1 & T_1 \\ T_1 & |T|^2 \end{array} \right) 
\;, \quad
G_{\M} = {1 \over T_2 V_{O_k}^i} \left( \begin{array}{cc} 1 & T_1 \\ T_1 & |T|^2 \end{array} \right).
\ee
For zero $B$-field, it is easy to see that these are in fact simply
\be
\Gamma_\A(t,T^i,a) &=& \sum_{m,n} e^{-\pi t (m^2 L_i^2 + n^2 D_i^2)} 
\ee
with the $L_i$ and $D_i$ from \eqref{LD}, 
and similarly for $\Gamma_{\M}$. 
We can now use the known result (see e.g.\ \cite{Berg:2004ek}) that
\be
\int_{1/\Lambda^2}^{\infty} \frac{dt}{t} \tht(itG)e^{-2\pi \chi t} &=& 
{\Lambda^2  \over \sqrt{G}} - \ln(8\pi^3 \chi) - \ln\left( {T_2 |\eta(T)|^4 \over \sqrt{G}}\right) \\
&=& V_a^i \Lambda^2 - \ln(8\pi^3 \chi) - \ln\left( {T_2 V_a^i|\eta(T)|^4}\right) \;. 
\ee

\subsection{Tadpole cancellation}
The charge cancellation condition is 
\be
\sum_a N_a(\Pi_a + \Pi_{a'})-2 \times 4\Pi_{O6} = 0 \; ,  \label{Picanc}
\ee
where the factor of $2$ arises from the fact that the O-planes come in pairs of two
in orientifolds with one rectangular torus (as is the case for our third torus). This can be seen
nicely in fig.\ 1 of \cite{Cvetic:2001nr}. 
For concreteness, we will focus on an example with two stacks $a$ and $b$,
with intersection numbers $(n_i,m_i)$ and $(q_i,p_i)$. 
Then \Ref{Picanc} becomes two conditions on the wrapping numbers:
\be \label{D1D2}
D_1 = 0 \; , \quad D_2 = 0\ ,
\ee
where
\be
D_1 &=& 
N_a(2n_1n_2 +n_1m_2+m_1n_2 -m_1m_2)n_3 \label{div1}  \\
&& +N_b(2q_1q_2
+q_1p_2+p_1q_2-p_1p_2)q_3 -8\ , \nonumber  \\ [2mm]
D_2 &=& 
-N_a(n_1m_2+m_1n_2+m_1m_2)m_3 -N_b (q_1p_2+p_1q_2+p_1p_2)p_3-8 \;. 
\label{div2}
\ee
We sketch the well-known derivation of this result, again specifying
to two stacks only (a generalization to more stacks is straightforward but a bit cumbersome to write down explicitly).
Take the UV limit in the vacuum amplitude, where
``UV" means in the open string sense $t\rightarrow 0$, that is $\ell\rightarrow \infty$,
and focus on the Ramond sector $(\alpha,\beta)=(1/2,0)$ piece
of the spin structure sum\footnote{This is sufficient for the vacuum amplitude as long as supersymmetry is not broken. In the presence of a background field $B$ supersymmetry is broken and, thus, for the $B^2$ terms, the NSNS tadpoles are independent of the RR tadpoles, cf.\  \cite{Lust:2003ky}.} 
and use that in this limit, 
\be
{\thbw{1/2}{0}(0) \over \eta^3}  \rightarrow  2
\; , \quad
{\thbw{1/2}{0}(v) \over
\thbw{1/2}{1/2}(v)}  \rightarrow  -\cot(\pi v)  \; . 
\ee
The sum of the UV limits of the vacuum amplitudes in the R sector is then
\be
\delta_R &=& \left(2 \A^{\rm UV}_{[a][b]}+\A^{\rm UV}_{[a][a]}+\A^{\rm UV}_{[b][b]}+\M^{\rm UV}_{[a]}+\M^{\rm UV}_{[b]} + \K^{\rm UV}\right) \int_0^{\infty} \!\!d\ell  \  \qquad \mbox{(R sector)} \ ,
\ee
where
\be
\A^{\rm UV}_{[a][b]} &=&  {1 \over 6} N_a N_b \sum_{k,l=0}^{2} (V_{a^k, b^l} + V_{a^k,\R b^l}
+V_{\R a^k,b^l} + V_{\R a^k,\R b^l} )\ , \\
\A^{\rm UV}_{[a][a]} &=& {1 \over 6} N^2_a  \sum_{k,l=0}^{2} (V_{a^k,a^l} + V_{a^k,\R a^l}+
V_{\R a^k,a^l} + V_{\R a^k,\R a^l} )\ , \\
\A^{\rm UV}_{[b][b]}&=&  {1 \over 6} N^2_b  \sum_{k,l=0}^{2} (V_{b^k,b^l} + V_{b^k,\R b^l}+
V_{\R b^k,b^l} + V_{\R b^k,\R b^l} )\ , \\
\M^{\rm UV}_{[a]}&=& {1 \over 3} N_a \sum_{k=0}^{2} \sum_{m=0}^{5}  (V_{a^k,O_m} + V_{\R a^k,O_m})\ , \\
 \M^{\rm UV}_{[b]} &=& {1 \over 3} N_b \sum_{k=0}^{2} \sum_{m=0}^{5}  (V_{b^k,O_m}+V_{\R b^k,O_m})\ ,  \\
\K^{\rm UV} &=&   {2 \over 3}\sum_{m,n=0}^{5} V_{O_m,O_n} \ . 
\ee
The additional factors of $2$ and $4$ in the M\"obius and Kleinbottle amplitudes have the same origin as the factor of $2$ in \eqref{Picanc}. Here we introduced the notation $V_{ab}=\prod_j V_{ab}^j$ (and similarly for $V_{aO}$ and $V_{OO}$) with each $V_{ab}^j$ given by \eqref{Vab}.\footnote{With a slight abuse of notation we use the same symbol $V_{ab}$ as in \eqref{Vab} to now denote $\prod_j V_{ab}^j$.} This can be obtained by first noticing 
\be
\A^{\rm UV}_{a,b} &=& 2 N_a N_b \prod_{j=1}^3 I^j_{ab} \cot \left(\pi v_{ab}^j\right) \int_0^{\infty} \!\!d\ell \ , \\
\M^{k,{\rm UV}}_{a, \R a^{ k}}  &=&  -8 N_a \rho_k \prod_{j=1}^3 I^{k;j}_{a,\R a^k} \cot \left(\pi v^j_{a, O_k} \right) \int_0^{\infty} \!\! d\ell \ , \\
\K^{\rm UV} &=& 16 \prod_{j=1}^3 I^{j}_{O_m O_n}  \int_0^{\infty} \!\! d\ell \ ,
\ee
where the factors $N$ and the phase $\rho_k$ come from the Chan-Paton traces of eq.\ (2.11) in \cite{Lust:2003ky}: 
\be \label{CPphase}
\tr ((\gamma_{\Omega \R \Theta^k}^{\Omega \R \Theta^k a})^*\gamma^a_{\Omega \R \Theta^k}) = \rho_k N_a \; .
\ee
We find that tadpoles cancel if $\rho_k=1$ for all $k$. 
One then uses \Ref{cotv} from above:
\be
\cot(v^j_{ab}) = {V^j_{ab} \over I^j_{ab}} \; ,
\ee
and similarly for $\cot(\pi v^j_{a, O_k})$.
We see that the intersection numbers in the angles
cancel the explicit overall intersection numbers from
multiple intersections in the amplitude. 

Demanding untwisted RR tadpole cancellation $\delta_R=0$ leads to \eqref{D1D2} which is a condition on $N_a$, $N_b$ and the wrapping numbers.


\subsection{Sample configuration}

This two-stack configuration of D6-branes 
(with the O6-plane configuration given above) cancels all untwisted tadpoles:
\be
a &=& [1,0,1,0,1,0] \\
b &=&  [0,1,0,1,0,-1]
\ee
and substituting this in \eqref{D1D2} gives the two constraints
\be
4N_a^2-32N_a+64=0 \; , \quad
3 N_b^2-48N_b+192=0 \; ,
\ee
whence $N_a=4$, $N_b=8$. 

\subsection{Divergence cancellation in the two-point function}
\label{divcancel}

Using the notation
of the previous sections, we can now write out
the divergences in terms of wrapping numbers
and $R_2/R_1$. 
There are no contributions from $\A_{[b][b]}$, $\M_{[b]}$ (for $b\neq a$) or $\K$ when calculating a 2-point function for the brane scalars of stack $a$. 
\be 
\langle
\Phi \bar \Phi \rangle_{\A_{[a][b]}}^{\rm UV} &=&  - 2N_aN_b\,  \Big(m_1(n_2-m_2)+n_1(2n_2+m_2)\Big) n_3 \Big(p_1(q_2-p_2)+q_1 (2q_2+p_2)\Big) q_3\,  {R_1 \over R_2} \nonumber \\
&& -6N_aN_b  \Big(m_1(n_2+m_2)+n_1m_2  \Big) m_3  \Big(p_1(q_2+p_2)+q_1p_2 \Big) p_3
\,  {R_2 \over R_1} \ , \\
\langle
\Phi \bar \Phi \rangle_{\A_{[a][a]}}^{\rm UV} &=&
\left[-2N_a^2 
 \Big(m_1(n_2-m_2)+n_1(m_2+2n_2) \Big)^2 n_3^2 \right]
  {R_1 \over R_2} \nonumber  \\
&& + \left[
-6N_a^2\Big(m_1(n_2+m_2)+n_1m_2\Big)^2m_3^2  \right] {R_2 \over R_1} \ , \\
\langle
\Phi \bar \Phi \rangle_{\M_{[a]} }^{\rm UV} 
&=&
\left[ 16N_a \Big(m_1(n_2-m_2)+n_1(2n_2+m_2)\Big)n_3
 \right] {R_1 \over R_2}
\nonumber  \\
&& + \left[ -48N_a \Big(m_1(n_2+m_2)+n_1m_2\Big)m_3
 \right] {R_2 \over R_1}\ .
\ee
We demand cancellation of
\be
\delta_{\rm 2pt} &=& \langle \Phi \bar \Phi \rangle_{\A_{[a][b]}}^{\rm UV} + \langle
\Phi \bar \Phi \rangle_{\A_{[a][a]}}^{\rm UV} +\langle
\Phi \bar \Phi \rangle_{\M_{[a]} }^{\rm UV} 
\ee
for {\it any} values of complex structures, by the prefactor of $ {R_1 \over R_2}$ and ${R_2 \over R_1}$ va\-ni\-shing. 
The result is two conditions for twelve integers. However,
this cancellation condition should not restrict the brane configurations any more 
than they have already been restricted by tadpole cancellation.
We find that as expected, the divergence factorizes
\be
\delta_{\rm 2pt} &=&
P_1 D_1  {R_1 \over R_2} + P_2 D_2  {R_2 \over R_1} \\
&=&0 \; ,  
\ee
where $P_1, P_2$ are cubic polynomials in the wrapping numbers, 
and $D_1$ and $D_2$ are the vacuum amplitude
tadpole cancellation conditions given in \Ref{div1} and \Ref{div2}. 
To be explicit, we find
\be
P_1 &=& 2 N_a(m_1 m_2 n_3 - n_2 n_3 m_1 - 2 n_1 n_2 n_3 - n_1 n_3 m_2) \ , \\ 
P_2 &=& 6 N_a (m_2 m_3 n_1  + n_2 m_3 m_1 + m_1 m_2 m_3) \; . 
\ee
Thus, the tadpole cancellation conditions $D_1=D_2=0$, that we compute by factorization of vacuum amplitudes, already imply
divergence cancellation in the scalar two-point function.

We have already imposed twisted tadpole cancellation by eq.\ (2.7) of \cite{Lust:2003ky} so we only see
the untwisted ones, cf.\ eq.\ (2.7) of \cite{Gmeiner:2007zz}. 

Finally, we consider the ``$I^3$'' contribution, where the integrand does not depend on the wrapping numbers at all. For these coefficients we find 
\be 
\sum_{k.l}I_{a^k, b^l} &=& - \sum_{k,l} I_{a^k, \R b^l}\ , \\
\sum_{k.l}I_{a^k, a^l}&=& \sum_{k,l} I_{a^k, \R a^l}  = 0\ , \\
\sum_{k.l}I_{a^k,O_l }  &=&  - \sum_{k.l}I_{\R a^k, O_l} \ , \label{Irelations}
\ee
so the total contribution vanishes.


\section{World-sheet correlators}
\label{wscorr}
The correlators on the annulus ${\cal A}$ and M\"obius strip ${\cal
M}$ are obtained by symmetrizing the 
corresponding correlators on the covering torus
under the involutions that define the surfaces in the first place:
\beqn \label{app:is}
I_{{\cal A}}(w) = I_{{\cal M}}(w) = 2\pi - {\bar w} 
\eeqn
producing (cf.\ the appendix of \cite{ABFPT})
\beqn
\langle X(w_1) X(w_2)\rangle _{\sigma}
=\langle X(w_1) X(w_2)\rangle _{{\cal T}} +\langle X(w_1)
X(I(w_2))\rangle _{{\cal T}} \ ,
\eeqn
where $\sigma \in \{ \ca, \cm\}$. The formulas are somewhat simpler in the rescaled variable
\be \label{nuw}
\nu = {w \over 2\pi} \; , \qquad  {\rm Re}(\nu) \in [0,1] \; .
\ee
The bosonic correlation function on the torus ${\cal T}$ in the
untwisted directions is
\beqn \label{boscorr}
\langle X(\nu_1,\bar{\nu}_1) X(\nu_2,\bar{\nu}_2)\rangle _{{\cal T}} =
-\frac{\alpha'}{2} \ln \Big|
   \frac{\vartheta_1({\nu_1-\nu_2},\tau )}{\vartheta_1'(0,\tau)}  \Big|^2
 + \alpha'  \pi \frac{({\rm Im}(\nu_1-\nu_2))^2}{  \,{\rm Im}(\tau)}\ .
 \eeqn
We will also need  the S-transformed expression on the annulus, for which $\nu$ and $\tau$ are imaginary
\be \label{XXS}
\langle X(\nu,\bar{\nu}) X(0,0)\rangle _{{\cal A}} =- \alpha' \ln \Big|
   \frac{\tau \vartheta_1({\nu \over \tau},-{1 \over \tau} )}{\vartheta_1'(0,-{1 \over \tau})}  \Big|^2 \ .
\ee
Since the bosons in the amplitude we are interested in
are polarized in external directions only,
we will not need twisted boson correlation functions in this paper. 

For untwisted world-sheet
fermions in the even spin structures, the correlation function on
the torus is
\beqn \label{fercor}
G_F\zba{\alpha}{\beta}(\nu_1, \nu_2) \delta^{\mu \nu} \equiv \langle
\psi^\mu(\nu_1)\psi^\nu(\nu_2)\rangle ^{\alpha,\beta}_{{\cal
T}}&=&
\frac{\vartheta[{\alpha \atop \beta}]({\nu_1-\nu_2} ,\tau)
{\vartheta_1^\prime}(0,\tau)}
{\vartheta[{\alpha \atop \beta}](0,\tau)
\vartheta_1({\nu_1-\nu_2} ,\tau)}\;
\delta^{\mu \nu} \ .
\eeqn
Just as for bosons, fermion propagators for the remaining surfaces
can be determined from the torus propagators by the method of
images. The result (taken from the appendix of \cite{ABFPT}) is 
\beqn \label{correlators}
\langle \psi(\nu_1)\psi(\nu_2)\rangle^{\alpha, \beta}_{\sigma} &=& G_F\zba{\alpha}{\beta}(\nu_1,\nu_2)\; , \quad
\sigma \in \{ \ca, \cm\}\ .
\eeqn
%

In the following we will sketch the derivation using periodicity of the doubled fermionic fields.
On the covering torus, we have for a worldsheet fermion $\psi$ with spin structure $(\alpha,\beta)$
\be
\psi(w+2\pi,\tau)&=& -e^{2\pi i \alpha} \psi(w,\tau)\ , \\
\psi(w+2\pi \tau,\tau)&=& -e^{-2\pi i \beta} \psi(w,\tau)\ .
\ee
The signs are conventional: they are chosen such that $(\alpha,\beta)=(1/2,1/2)$
corresponds to double periodicity. 
Thus, for the Green's function we
are looking for an expression that transforms under translations around the two cycles
of the covering torus as
\be  \label{Gtrans_1}
G_F\zba{\alpha}{\beta}(w+2\pi,\tau)&=& -e^{2\pi i \alpha} G_F\zba{\alpha}{\beta}(w,\tau)\ , \\
\label{Gtrans_2}
G_F\zba{\alpha}{\beta}(w+2\pi \tau,\tau)&=& -e^{-2\pi  i\beta} G_F\zba{\alpha}{\beta}(w,\tau) 
\ee
and satisfies
\be
\bar{\partial} G_F\zba{\alpha}{\beta}(w,\tau) = \delta(w)    \; , \quad (\alpha,\beta) \neq (1/2,1/2) \; . 
\ee
This determines the expression to be
\be  \label{GFtheta}
G_F\zba{\alpha}{\beta}(\nu ,\tau) = \frac{\vartheta[{\alpha \atop {\beta}}](\nu) \tht_1'(0)}{\vartheta[{\alpha \atop {\beta}}](0)
\tht_1(\nu) }    \; , \quad (\alpha,\beta) \neq (1/2,1/2) \; ,
\ee
where we fixed the residue at $\nu=0$ to be 1. 
In this subsection, $\alpha$ has been a generic real number between 0 and 1. 
To connect to the discussion in the main text, we now give a concrete example for the annulus amplitude.
In that case, for angles $v=\varphi/\pi$
we have that the generic $\alpha$ above
is actually $\tilde{\alpha} + v$, where $\tilde{\alpha}$ is  now  0 or 1/2.
In the main text we drop the tilde and let $\alpha$ only take the values 0 or 1/2. 

By using modular transformations of the Jacobi theta functions,
it is easy to see that
\be \label{GFS}
G_F\zba{\alpha}{\beta} \left({\nu / \tau},-{1 / \tau}\right) = \tau G_F\zba{\beta}{-\alpha} (\nu,\tau)\ .
\ee


 \section{$q$-series representation of twisted correlator}
 \label{WW}
\begin{figure}[th]
\begin{center}
\includegraphics[scale=0.6]{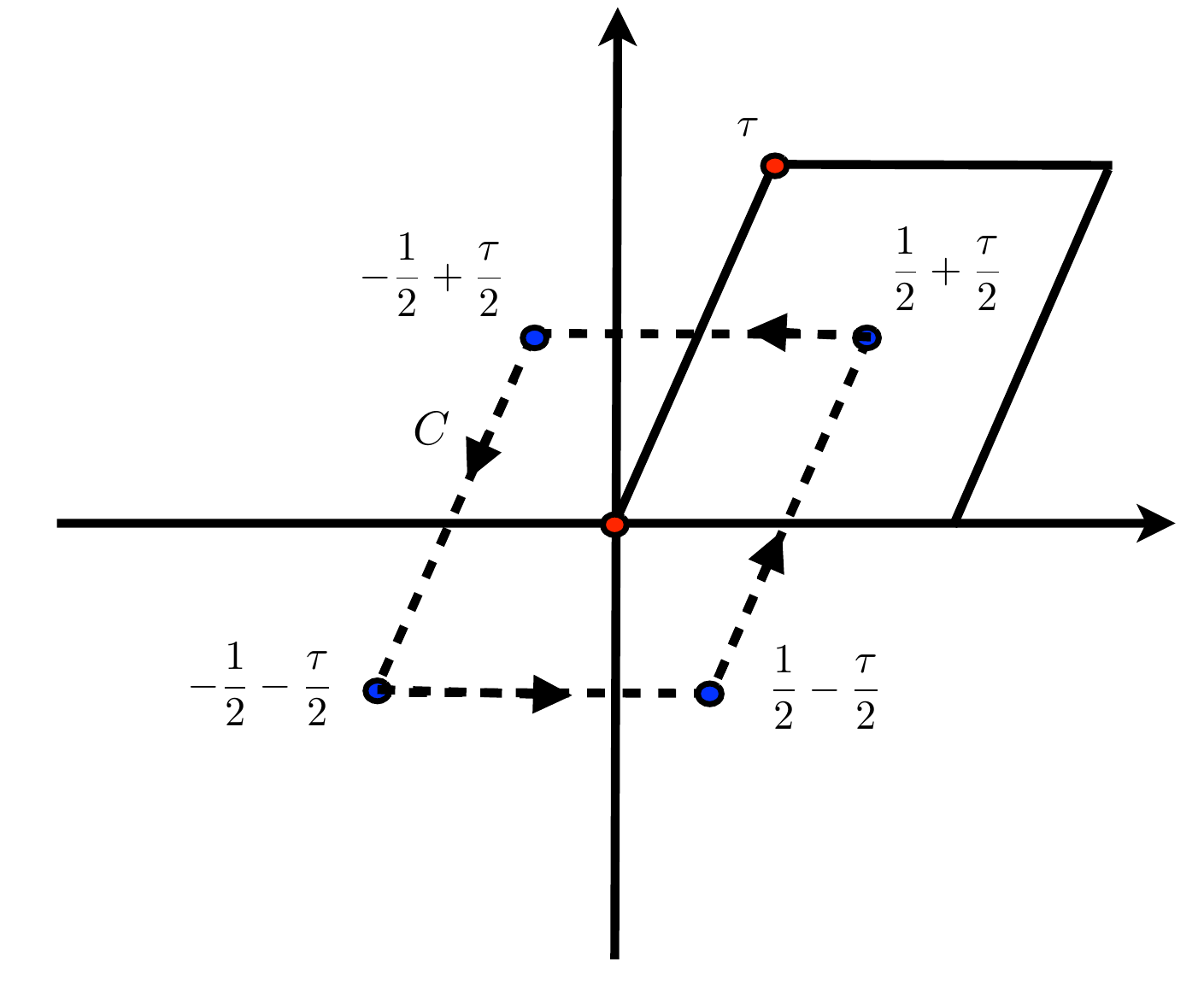}
\caption{Path of integration $C$ in the complex $z$ plane.}
\vspace{-2mm}
\label{cell}
\end{center}
\end{figure}

We want to find a different representation of \eqref{GFWW}. We begin by observing
\be
\vartheta\zba{1/2}{1/2 + v}(\nu,\tau) = \vartheta \zba{1/2}{1/2} (\nu+v,\tau)\ . 
\ee
as is obvious from the sum representation of the Jacobi theta function. 
Then we have
\be
G_F\zba{1/2}{1/2+v}(\nu ,\tau) = {\tht_1(\nu+v,\tau) \tht_1'(0,\tau) \over \tht_1(v,\tau)\tht_1(\nu,\tau)}  \; .
\ee
In the main text, the left hand side is the fermion Green's function in the closed string channel. 
In this section only, $\tau$ 
is not the specific open-string channel $\tau$ of the main text, but rather we will derive a general identity for generic $\tau$.
Also, for clarity we relabel $\nu=y$, $v=z$ in this section only. We will prove that
\be  \label{target}
f(y,z)\equiv {\tht_1(y+ z)\,\tht_1' (0) \over \tht_1(y) \tht_1(z)} =\pi \cot \pi y + \pi \cot \pi z + 4 \pi \sum_{m=1}^{\infty} \sum_{n=1}^{\infty} q^{mn} \sin(2 \pi m y + 2 \pi nz)
\ee
with $q = e^{2 \pi i \tau}$. This is literally a textbook problem, exercise
13 in Chapter 21 of \cite{WandW}. For the reader's convenience we will solve this problem here.
Note that $f(y,z)$ is symmetric under interchange $y \leftrightarrow z$. (This
is rather interesting in the original variables, as one would
in general not expect the integrand to be symmetric in the vertex position and the angle.)
We will concentrate on the $z$ dependence of $f(y,z)$, assuming that $y$ is away from zero.
To prove \eqref{target}, perform the contour integration around the
cell in figure \ref{cell}.
\be
{1 \over 2\pi i}
\oint_C {f(y,z)} e^{2\pi i n z} dz
&=& {1 \over 2\pi i} \left[
\int_{-1/2-\tau /2}^{1/2-\tau /2} dz \left( f(y,z) e^{2\pi i n z} -
{f(y, z+\tau)} e^{2\pi i n (z+\tau)} \right)
\right]   \label{fint} \\
&&
-{1 \over 2\pi i} \left[
\int_{-1/2-\tau /2}^{-1/2+\tau/2}dz\left( f(y,z) e^{2\pi i n z} -
f(y, z+1) e^{2\pi in (z+1)} \right)
\right]  \nonumber \\
&=& {1 \over 2\pi i} \left[
\int_{-1/2-\tau /2}^{1/2-\tau /2}dz\left( f(y,z) (e^{2\pi i n z} -
 e^{2\pi in (z+\tau)-2\pi iy})\right)
\right] \nonumber \\
&&
-{1 \over 2\pi i} \left[
\int_{-1/2-\tau /2}^{-1/2+\tau /2}dz\left( f(y,z) (e^{2\pi i n z} -
 e^{2\pi in (z+1)}) \right) 
\right]  \nonumber  \\
&=&
{1 \over 2\pi i} (1-q^{n}e^{-2\pi i y})\int_{-1/2-\tau /2}^{1/2-\tau/2}dz  f(y,z) e^{2\pi i n z}\ ,
\label{final_1}
\ee
where $n$ is positive integer and we used $f(y,z+1)=f(y,z)$ and $f(y,z+\tau)=e^{-2 \pi i y} f(y,z)$, which follow from the properties of theta functions,
or more directly, from the defining transformation properties
\eqref{Gtrans_1}, \eqref{Gtrans_2}. In fact, we  did not even need to write down the theta function representation
of $f$ to 
complete the proof, we only need to know its quasiperiodicity properties and singularity (including the residue). Knowing that $f$ has a simple pole
at the origin with residue one lets us immediately see that the integral
on the left-hand side of \eqref{fint} is equal to one, because
\be 
{\rm Res}[f(y,z) e^{2 \pi i n z}, z=0] = 2 \pi i\ .
\ee
Thus, we obtain
\be 
\int_{-1/2-\tau /2}^{1/2-\tau/2}dz f(y,z)  e^{2\pi i n z} = { 2\pi i \over 1-q^{n}e^{-2\pi i y}}\ ,
\ee 
and assuming $|q^n e^{-2 \pi i y}| <1$, the RHS can be Taylor expanded to give
\be \label{WW13}
\int_{-1/2-\tau /2}^{1/2-\tau/2}dz f(y,z)  e^{2\pi i n z} = 2\pi i \sum_{m=0}^{\infty} q^{m n}e^{-2\pi i m y}\ .
\ee
On the other hand, from the pole structure and the periodicity of $f(y,z)$ we can use the following ansatz:
\be  \label{ansatz}
f(y,z)=\pi \cot (\pi y) + \pi \cot (\pi z) + \sum_{m=-\infty}^{\infty}\sum_{n=-\infty}^{\infty} c_{m,n} e^{-2 \pi i n z} e^{-2 \pi i m y},
\ee 
where the cotangent terms arise due to the fact that the pole is located at zero. 
(The Fourier series of the cotangent function itself are written down in section \ref{cotfourier} below.)
And note that $f(-y,-z)=-f(y,z)$, so it is easy to see that
\be
c_{m,n}=-c_{-m,-n} \label{odd},
\ee
meaning in particular $c_{0,0}=0$. On the other hand due to the symmetry under $y \leftrightarrow z$ it is easy to show 
\be \label{symmetry}
c_{m,n}=c_{n,m}.
\ee
Inserting these pieces of information into the integral in (\ref{WW13}) and using the expansion
(\ref{cot}) below for $\pi \cot(\pi z)$, we find
\be
2 \pi i + \sum_{m=-\infty}^{\infty} c_{m,n} e^{-2 \pi i m y}= 2\pi i \sum_{m=0}^{\infty} q^{m n}e^{-2\pi i m y},
\ee
which by matching term by term gives
\be
c_{m,n}= 2 \pi i q^{mn} \;, \quad c_{-m,n} = 0 \; , \quad c_{0, n} = 0 \;  \qquad \mbox{for $m, n>0$}\ ,
\ee
and using (\ref{odd}) we have
\be
c_{-m,-n}= - 2\pi i q^{mn} \;, \quad c_{m,-n} = 0 \; , \quad c_{0, -n} = 0 \;  \qquad \mbox{for $m, n>0$}\ ,
\ee
and due to (\ref{symmetry}) it follows that $c_{m,0}=0$ for any $m$.

To summarize, we are left with
\be 
\sum_{m=-\infty}^{\infty}\sum_{n=-\infty}^{\infty} c_{m,n} e^{-2 \pi i n z} e^{-2 \pi i m y} &=& 
2 \pi i \sum_{m=1}^{\infty}\sum_{n=1}^{\infty} q^{mn} e^{-2 \pi i n z} e^{-2 \pi i m y} \\ \nonumber
&&- 2\pi i \sum_{m=-1}^{-\infty}\sum_{n=-1}^{-\infty} q^{mn} e^{-2 \pi i n z} e^{-2 \pi i m y} \\ \nonumber
&=& 2 \pi i \sum_{m=1}^{\infty}\sum_{n=1}^{\infty} q^{mn} \left(e^{-2 \pi i n z} e^{-2 \pi i m y}-e^{2 \pi i n z} e^{2 \pi i m y}\right) \\ \nonumber
&=& 4 \pi \sum_{m=1}^{\infty}\sum_{n=1}^{\infty} q^{mn} \sin(2 \pi m y +2 \pi nz). \\ \nonumber
\ee
Thus (\ref{ansatz}) gives (\ref{target}), which is what we wanted to show. 


\subsection{Vanishing by contour integration}
\label {app:vanishing}
\begin{figure}[th]
\begin{center}
\includegraphics[scale=0.7]{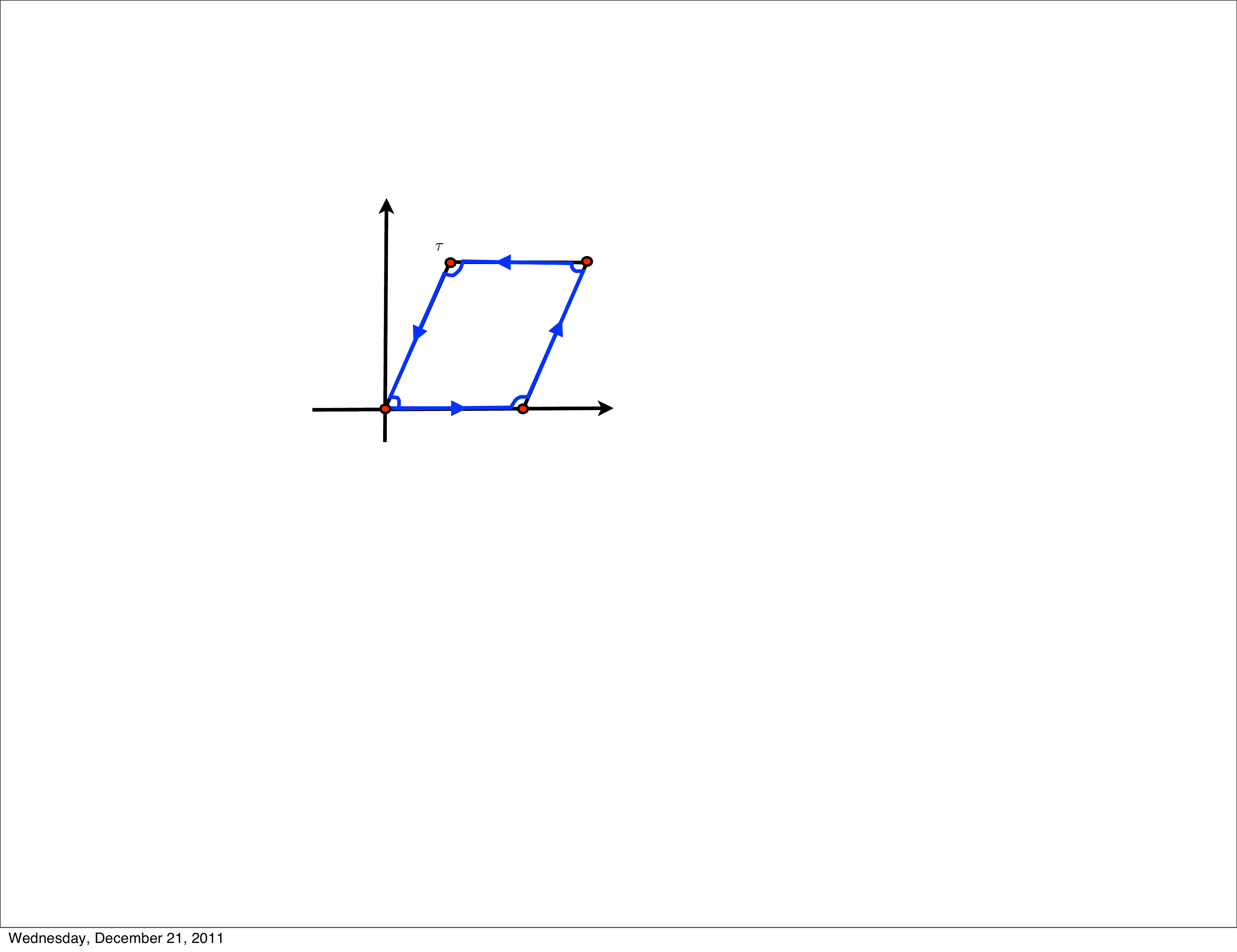}
\caption{Path of integration in the complex $\nu$ plane.}
\vspace{-2mm}
\label{cellGF}
\end{center}
\end{figure}

Once we have convinced ourselves
that the poles do not contribute, we can prove \eqref{vanishing} by 
performing a line integral over the deformed contour of figure \ref{cellGF}. 
Because $G_F$ has no pole in the interior of the fundamental domain, we find
by a similar argument to above
\be
0 = \oint d\nu \;  G_F\zba{1/2}{1/2+v}(\nu) = (1-e^{-2\pi i v}) \int_0^1 d\nu \;G_F\zba{1/2}{1/2+v}(\nu) \; .
\ee
If $e^{2\pi i v} \neq 1$, the factor in parenthesis does not vanish, then the integral along the real axis
from 0 to 1 of $G_F$ must instead vanish, which is what we wanted to show. 
We note that if it had not been for the quasiperiodicity induced by 
the angle $v$, the factor in parenthesis would vanish trivially, and the contour integration would
 provide no information about the value of the integral  of $G_F$ from 0 to 1. 


\subsection{Fourier series of cotangent function}
\label{cotfourier}
For $|e^{2\pi i \nu}|<1$ we have that (writing $x=e^{2\pi i \nu}$)
\be
\label{Apostolcot}
\pi \cot \pi \nu &=& \pi i {e^{2\pi i \nu}+1 \over e^{2\pi i \nu}-1}
= -\pi i \left({x \over 1-x} + {1 \over 1-x}\right)\\
&=&  -\pi i \left(\sum_{n=1}^{\infty}x^n +\sum_{n=0}^{\infty}x^n \right)
= -\pi i \left( 1 + 2 \sum_{n=1}^{\infty}e^{2\pi i n\nu}\right)\ ,  \nonumber
\ee
but for $|e^{-2\pi i \nu}|<1$ we have that
\be \label{cot}
\pi \cot \pi \nu &=& \pi i {1+ e^{-2\pi i \nu} \over 1-e^{-2\pi i \nu}}
= +\pi i \left({1 \over 1-1/x} + {1/x \over 1-1/x}\right)\\
&=&  +\pi i \left(\sum_{n=1}^{\infty}\left(1 \over x\right)^n+\sum_{n=0}^{\infty}\left(1 \over x\right)^n  \right)
= +\pi i \left( 1 + 2 \sum_{n=1}^{\infty}e^{-2\pi i n\nu}\right)\ .  \nonumber
\ee


\section{Illustrating image intersections}
\label{app:invint}

\begin{figure}
\begin{center}
\includegraphics[width=0.6\textwidth]{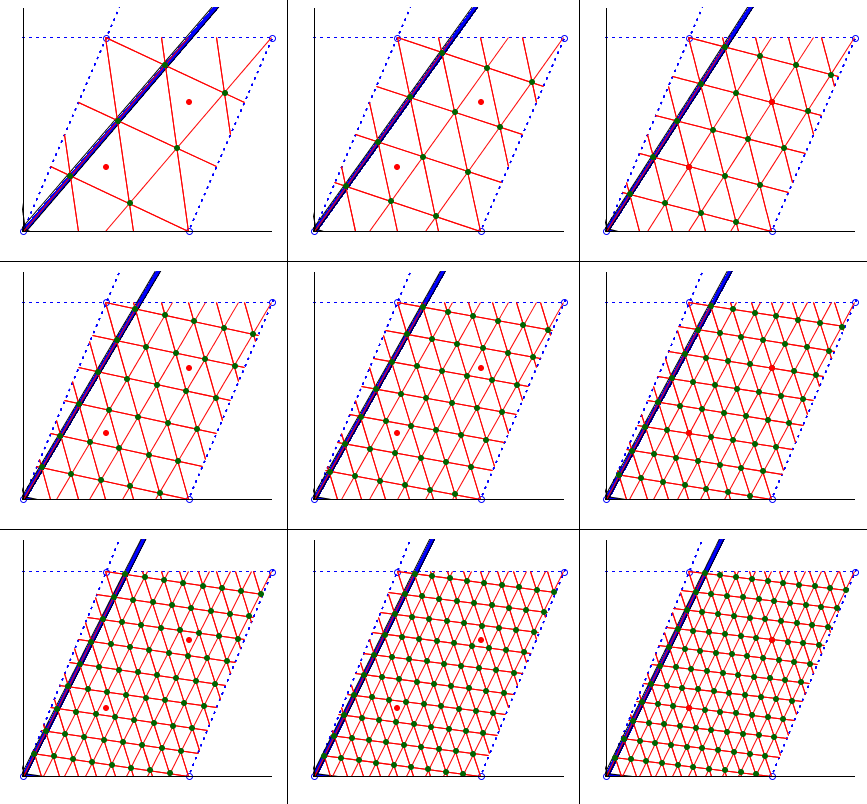}
\caption{
Orbifold invariant intersection points for $[a]$ given by $(n,m)=(n,1), n=2, \ldots, 10$.
We see that including the origin, $I_{[a][a]} = 7,13,21,31,43, \ldots  = n^2 + n  +1 $.}
\label{braneimage}
\end{center}
\end{figure}

Let us focus on a single brane $a$ and its images $a_k$ on a single $T^2$
let us say the second one. 
From \eqref{wrapZ6} we have
\be
I_{[a][a]}&=& I_{a, \Theta a} +  I_{a, \Theta^2 a} \\
&=& {1 \over 2}\left[ n_1\cdot (n_1+m_1)+m_1\cdot m_1 \right]
+{1 \over 2} \left[ n_1\cdot n_1 + m_1 \cdot (n_1+m_1)\right] \\
&=& n_1^2 + n_1 m_1 + m_1^2 \ . 
\ee
For example, for $[a]$ given by $(n_1,m_1)=(n,1)$ we have for the orbit that
\be
I_{[a][a]} &=& n^2+n+1\\
&=& 7,13, 21, 31, 43, \ldots \; . 
\ee
This is illustrated in  figure \ref{braneimage}. 

\end{document}